\documentclass[fleqn,usenatbib]{mnras}

\usepackage{newtxtext,newtxmath}

\usepackage[T1]{fontenc}

\DeclareRobustCommand{\VAN}[3]{#2}
\let\VANthebibliography\thebibliography
\def\thebibliography{\DeclareRobustCommand{\VAN}[3]{##3}\VANthebibliography}

\usepackage{cleveref}
\usepackage{caption}
\usepackage{soul}
\usepackage{siunitx}
\usepackage[dvipsnames]{xcolor}
\usepackage{graphicx}	
\usepackage{amsmath}	

\DeclareSIUnit\angstrom{\text {Å}}

\title[GAMA/DEVILS: Cosmic star formation and AGN activity]{GAMA/DEVILS: Cosmic star formation and AGN activity over 12.5 billion years}

\author[J. C. J. D'Silva et al.]
{Jordan C. J. D'Silva,$^{1,2}$\thanks{E-mail: jordan.dsilva@research.uwa.edu.au}
Simon P. Driver,$^{1}$
Claudia D. P. Lagos,$^{1,2}$
Aaron S. G. Robotham,$^{1,2}$  \newauthor
Sabine Bellstedt,$^{1}$
Luke J. M. Davies,$^{1}$
Jessica E. Thorne,$^{1}$ 
Joss Bland-Hawthorn,$^{3,2}$
Matias Bravo,$^{4}$ \newauthor
Benne Holwerda,$^{5}$
Steven Phillipps,$^{6}$ 
Nick Seymour,$^{7}$ 
Malgorzata Siudek,$^{8,9}$ 
Rogier A. Windhorst,$^{10}$
\\
$^{1}$ICRAR, The University of Western Australia, 35 Stirling Highway, Crawley, WA 6009, Australia\\
$^{2}$ARC Centre of Excellence for All Sky Astrophysics in 3 Dimensions (ASTRO 3D)\\
$^{3}$Sydney Institute for Astronomy, School of Physics, A28, The University of Sydney, NSW 2006, Australia\\
$^{4}$Department of Physics \& Astronomy, McMaster University, 1280 Main Street W, Hamilton, ON, L8S 4M1, Canada\\
$^{5}$University of Louisville, Department of Physics and Astronomy, 102 Natural Science Building, 40292 KY Louisville, USA\\
$^{6}$Astrophysics Group, School of Physics, University of Bristol, Tyndall Avenue, Bristol BS8 1TL, UK\\
$^{7}$ICRAR, Curtin University, Bentley, WA 6102, Australia\\
$^{8}$Institute of Space Sciences (ICE, CSIC), Campus UAB, Carrerde Can Magrans, s/n, 08193 Barcelona, Spain\\
$^{9}$Institut de Física d’Altes Energies (IFAE), The Barcelona Institute of Science and Technology, 08193 Bellaterra (Barcelona), Spain\\
$^{10}$School of Earth and Space Exploration, Arizona State University, Tempe, AZ 85287-1404, USA}

\date{Accepted XXX. Received YYY; in original form ZZZ}

\pubyear{2022}

\begin{document}
\label{firstpage}
\pagerange{\pageref{firstpage}--\pageref{lastpage}}
\maketitle

\begin{abstract}
We use the Galaxy and Mass Assembly (GAMA) and the Deep Extragalactic Visible Legacy Survey (DEVILS) observational data sets to calculate the cosmic star formation rate (SFR) and active galactic nuclei (AGN) bolometric luminosity history (CSFH/CAGNH) over the last 12.5 billion years. SFRs and AGN bolometric luminosities were derived using the spectral energy distribution fitting code \textsc{ProSpect}, which includes an AGN prescription to self consistently model the contribution from both AGN and stellar emission to the observed rest-frame ultra-violet to far-infrared photometry. We find that both the CSFH and CAGNH evolve similarly, rising in the early Universe up to a peak at look-back time $\approx 10$~Gyr ($z \approx 2$), before declining toward the present day. The key result of this work is that we find the ratio of CAGNH to CSFH has been flat ($\approx 10^{42.5}\mathrm{erg \, s^{-1}M_{\odot}^{-1}yr}$) for $11$~Gyr up to the present day, indicating that star formation and AGN activity have been coeval over this time period. We find that the stellar masses of the galaxies that contribute most to the CSFH and CAGNH are similar, implying a common cause, which is likely gas inflow. The depletion of the gas supply suppresses cosmic star formation and AGN activity equivalently to ensure that they have experienced similar declines over the last 10 Gyr. These results are an important milestone for reconciling the role of star formation and AGN activity in the life cycle of galaxies.
\end{abstract}




\section{Introduction}

The cosmic optical and infrared backgrounds (COB \& CIB) are the second largest repositories of photon energy in the Universe. Each is about a factor 40 lower than the cosmic microwave background (CMB) \citep[e.g.,][]{hill_2018_cosmic_background}. While the CMB is comprised of the relic photons produced during the Big Bang, the COB and CIB is produced from the galaxy population as it evolves \citep[e.g.,][]{driverMeasurementsExtragalacticBackground2016a} -- from star-formation and the accretion of matter onto central super-massive black holes. Hence the COB and CIB represents an encoded record of the star formation and active galactic nuclei (AGN) activity modulated by dust reprocessing as the light escapes the host galaxies. As such, if we have an accurate representation of these two ingredients, the cosmic star-formation history (CSFH) and the cosmic AGN bolometric luminosity history (CAGNH), the COB and CIB should be fully predictable. 

 The CSFH is the combined sum of star formation rate (SFR) per unit volume, or cosmic SFR density, as a function of time or redshift. The CSFH peaks at look-back time $\approx 10$~Gyr ($z\approx2\to3$) before steadily declining into the present day, indicating an initial rapid formation of stars in the first few billion years since the Big Bang to a relaxed phase of star formation thereafter \citep{madauCosmicStarFormationHistory2014}. A similar story is also seen in the evolution of the cosmic ultra violet (UV) luminosity density, i.e., the sum of UV luminosities per unit volume as a function of time or redshift \citep{lillyCanadaFranceRedshiftSurvey1996}, highlighting the link between UV light and star formation. 

Star formation activity is not only encoded in the UV light but in the entire spectral energy distributions (SEDs) of galaxies, from gamma rays through to radio wavelengths \citep[e.g.,][]{kennicuttStarFormationMilky2012,daviesGalaxyMassAssembly2017a,FermiLAT2018EBL}. For example, UV photons from recently formed stars can also heat up the surrounding dust medium so that the star formation activity can be estimated from the thermal far-infrared emission of the dust. However, the assumptions that go into converting observed light into SFR can lead to situations where no two indicators agree on the SFR of the same galaxies, which propagate through to inconsistencies in the shapes of relations that scale with SFR, as well as the CSFH \citep{daviesGAMAHATLASMetaanalysis2016a,casey_csfh_2018}. Thus, it is preferable to consider star formation not at isolated wavelengths but across the full breadth of the SEDs of galaxies. This is essentially what SED fitting does: to derive physical quantities of galaxies by considering most, if not all, of the processes that produced the galaxy's SED \citep{conroyModelingPanchromaticSpectral2013}. 

However, as galaxies are multi-component in nature, the contribution of flux to their SEDs is not only from stellar emission. AGN, the central engines of galaxies, can have appreciable effects on the properties derived from fitting galaxy SEDs, which leads to uncertainties in those derived quantities. For example, SFRs derived from SED fitting can decrease by as much as $2$~dex when including an AGN component because erroneously allocated star formation flux is more correctly assigned to the AGN component \citep{thorneDeepExtragalacticVIsible2022a}. For this reason, galaxies with strong AGN dominated SEDs are often excluded in the analysis of the CSFH to avoid AGN contamination biasing the interpretation of the results.

One issue with this approach is that not all AGN contribute similarly to their galaxy's SED, and so the efficiency at which they are detected, and thus filtered out, varies from wavelength to wavelength as a result of the underlying physics of AGN (e.g., the presence of radio jets, dust geometry, environment) and selection effects between different wavelength observatories \citep[e.g.,][]{padovani_agn_2017}. In their determination of the CSFH using the Galaxy and Mass Assembly (GAMA), G10-COSMOS and 3D-HST data sets, \citet{driverGAMAG10COSMOS3DHST2018} filtered AGN with particular selection cuts that relied on a combination of the mid-IR emission \citep[e.g.,][]{Donley2012AGN}, radio emission \citep{seymour_star_2008} and X-ray emission \citep{laigle_cosmos2015_2016}. \citet{driverGAMAG10COSMOS3DHST2018} varied the leniency of these AGN cuts to estimate a maximum uncertainty on the CSFH induced by AGN selection of $\approx 0.1$ dex at look-back time $>10$~Gyr and $\lesssim 0.03$~dex uncertainty for look-back time $<10$ Gyr. 

While it is encouraging that this approach does not induce a significant error upon the CSFH, the choice to exclude AGN means that we can never reconcile the union of star formation and AGN processes into the broader context of galaxy formation. The appreciable effects that AGN have upon the SEDs is tantamount to AGN having an important role in shaping the physics of the host galaxies. For example, the presence of AGN is thought to affect the correlation between stellar mass and SFR, known as the star forming main sequence \citep{brinchmannPhysicalPropertiesStarforming2004,noeskeStarFormationAEGIS2007,salimUVStarFormation2007,Whitaker2012sfs}, by injecting energy into galaxies and inhibiting star formation \citep{katsianis_ssfr_19,matthee_scatter_19,davies_sigmassfr_19,daviesDeepExtragalacticVIsible2022}. Furthermore, AGN and star formation are more directly linked by their common correlations with galaxy properties like stellar mass and black hole velocity dispersion \citep[e.g.,][]{faberVelocityDispersionsMasstolight1976,ferrareseFundamentalRelationSupermassive2000}. These correlations emphasise the need to understand the role AGN play in the grand scheme of galaxy formation and evolution. Thus, a complete picture of galaxy formation requires a parallel view of both star formation and AGN activity as opposed to an either/or scenario, and hence the need to actually disentangle the light from stars and AGN components.

In this work, we aim to quantify the statistical evolution of star formation and AGN activity over $\approx 12.5$~Gyr from the present day. We expand upon the work of \citet{driverGAMAG10COSMOS3DHST2018,thorneDeepExtragalacticVIsible2021a,thorneDeepExtragalacticVIsible2022a} and investigate this evolution through the lenses of the CSFH and the cosmic AGN bolometric luminosity history (CAGNH), for which AGN and star formation flux have been self-consistently accounted. To do this, we use a combination of GAMA and Deep Extragalactic VIsible Legacy Survey (DEVILS) observations whose SFRs and AGN bolometric luminosities have been derived using the SED fitting code \textsc{ProSpect} \citep{robothamProSpectGeneratingSpectral2020}. The result is a quantitative depiction of star formation and AGN activity over $\approx 12.5$ Gyr of cosmic time. In \cref{sect:data}, we detail the data sets that were used in this work. In \cref{sect:methods} we detail our methods for deriving the CSFH and CAGNH. In \cref{sect:results-and-discussion}, we present our main results and discussions, and in \cref{sect:conclusion} we present our conclusions. Unless otherwise specified, we use the AB magnitude system \citep{okeSecondaryStandardStars1983}, the Chabrier initial mass function \citep[IMF,][]{chabrier_galactic_2003}, and standard `concordance' cosmology ($H_{0} = 70\; \mathrm{km s^{-1}Mpc^{-1}}$, $\Omega_{M} = 0.3$, $\Omega_{\Lambda} = 0.7$).

\section{Data}
\label{sect:data}

\subsection{GAMA}
\label{subsect:gama_devils}
GAMA is a wide area spectroscopic survey across five fields (G02, G09, G12, G15, G23) conducted on the Anglo-Australian Telescope (AAT) that has secured the redshifts of $\sim 300 \, 000$ galaxies \citep{driverGalaxyMassAssembly2011,liskeGalaxyMassAssembly2015,baldryGalaxyMassAssembly2018}. The four primary GAMA fields (G09, G12, G15 and G23) together have a spectroscopic redshift completeness of 95 percent to a limiting magnitude of 19.65 in the \textit{r}-band \citep{driver_galaxy_2022}. The survey is also complemented by panchromatic imaging of 20 bands from the far ultraviolet to the far infrared \citep{driverGalaxyMassAssembly2016a}, amassed from a compilation of GALEX \textit{FUV} and \textit{NUV} \citep{zamojski_deep_2007}; VST \textit{u, g, r, i} \citep{kuijken_kids_2019}; VISTA \textit{Z, Y, J, H, K$_{\mathrm{s}}$} \citep{mccracken_ultravista_2012}; WISE \textit{W1, W2, W3, W4} \citep{wright_wise_2010} and \textit{Herschel P100, P160, S250, S350, S500} \citep{lutz_pacs_2011,hermes_collaboration_herschel_2012} observations. 

\citet{bellstedt_profound_2020} used the source detection software \textsc{ProFound} \citep{robothamProFoundSourceExtraction2018} on a combined stack of the r+Z band images to compute matched segment photometry for the FUV-FIR.\textsc{ProFound} implements an iterative deblending algorithm where sources are thought of as troughs of flux in the images. This is referred to as `watershed deblending' in \citet{robothamProFoundSourceExtraction2018}. Manual reconstruction of spuriously fragmented segments was required for less than 1 per cent of the entire sample. \textsc{ProFound} was then run in multiband mode to process photometry of all bands from the FUV-FIR. 
After artefact and star flagging, the total effective area sampled in the \textsc{ProFound} GAMA catalogue is $\approx 217.54$ deg$^{2}$. \citet{bellstedt_prospect_2020} used the SED fitting code \textsc{ProSpect} \citep{robothamProSpectGeneratingSpectral2020} to produce a catalogue of physical quantities, such as stellar masses and SFRs, for $\sim 230\,000$ GAMA galaxies, which had secured redshifts $z>0$ with a data quality flag $nQ>3$ and were also securely classed as galaxies in the photometric catalogue. As \citet{bellstedt_prospect_2020} only fit the SEDs of GAMA galaxies with a stellar component, \citet{thorneDeepExtragalacticVIsible2022a} revisited the SED fits and used \textsc{ProSpect} with the inclusion of an AGN component to produce a catalogue of not only stellar masses and SFRs but AGN bolometric luminosity also. We use both catalogues of physical quantities in this work. 

\subsection{DEVILS-D10}
DEVILS \citep{daviesDeepExtragalacticVIsible2018a} is the companion survey to GAMA. Like GAMA, DEVILS is a spectroscopic survey conducted on the AAT covering $\approx 4.5$ deg$^{2}$ 
within three well studied fields (D02 $\to$ XMM-LSS, D03 $\to$ ECDFS, D10 $\to$ COSMOS) to a limiting magnitude of Y$<20/19.6/20.7$ mag in each of the three fields \citep[][ Davies et al. in prep]{daviesDeepExtragalacticVIsible2018a}.

We choose to focus only on the $\approx 1.5$ deg$^{2}$ DEVILS-D10 field as it has the highest spectroscopic redshift completeness (>85 per cent) of the three fields, being complemented by the previous spectroscopic campaign $z$COSMOS \citep{lillyZCOSMOS10kBrightSpectroscopic2009}. The DEVILS-D10 field is also prioritised due to the high quality photometric redshifts provided from COSMOS2015 \citep{laigle_cosmos2015_2016}.
The photometric catalogue for the DEVILS-D10 sub-survey consists of 22 
band imaging from GALEX \textit{FUV} and \textit{NUV}; CFHT \textit{u} \citep{capak_first_2007}; HSC \textit{g, r, i, z} \citep{aihara_second_2019}; VISTA \textit{Y, J, H, K$_{\mathrm{s}}$}; \textit{Spitzer IRAC1, IRAC2, IRAC3, IRAC4, MIPS24, MIPS70} \citep{laigle_cosmos2015_2016,sanders_spitzer_2007} and \textit{Herschel P100, P160, S250, S350, S500} observations (see \citealt{davies_deep_2021} for further details).

\textsc{ProFound} was used for source extraction and measurement in a two-phased approach, where the UV-MIR and MIR-FIR were processed separately. For the UV-MIR photometry, detection was performed on an inverse variance weighted stack of the UltraVISTA Y, J and H images. Spurious fragmentation and over-blending, which especially affects crowded areas of the detection images, were manually reconstructed. The corrected segmentation map was then used for multiband processing on the GALEX \textit{FUV} to \textit{Spitzer IRAC} 8$\mu$m images. For the MIR-FIR bands (24$\mu$m $\to$ 504$\mu$m), photometry was measured by using a Bayesian point spread function fitting (PSF) approach where the source locations from the initial \textsc{ProFound} run were used as spatial inputs on this longer wavelength imaging and the flux was extracted from the fitted PSF.

We note that the PSF fitting approach was only used on galaxies with $Y<21$ mag meaning that some galaxies are missing FIR photometry (the fluxes and flux errors are set to NA). This directly affects our work because \citet{thorneDeepExtragalacticVIsible2022a} found that FIR photometry is critical in constraining the AGN component of the SEDs of DEVILS galaxies. In DEVILS-D10, the percentage of Y $< 21.2$ mag galaxies with <30 mag FIR counterparts varies from 81.1 per cent at 24$\mu$m to 28.3 per cent at 504$\mu$m \citep{davies_deep_2021}. We note that the measured depths of the $24 \to 504 \mu$m photometry vary between $\approx 18.5 \to 14.0$ mag (see table 1 in \citet{davies_deep_2021} for the full comparison of measured depths in DEVILS-D10). A similar PSF fitting approach was employed for the FIR bands in GAMA; though, all optically detected objects were registered to have FIR counterpart fluxes, and so the effect on the AGN component is minimal. 


After star and artefact masking, the effective area covered by DEVILS-D10 is 1.47 deg$^{2}$. The result is $\sim 500 \, 000$ galaxies with $\approx$1, 5, 94 per cent having grism, spectroscopic or photometric redshifts. Catalogues of physical quantities of DEVILS galaxies were first presented in \citet{thorneDeepExtragalacticVIsible2021a} who used \textsc{ProSpect} to fit the galaxy SEDs. \citet{thorneDeepExtragalacticVIsible2021a} fit the SEDs of DEVILS-D10 galaxies with a stellar component only, much like in the case of GAMA. \citet{thorneDeepExtragalacticVIsible2022a} refitted the DEVILS-D10 SEDs using \textsc{ProSpect} with the inclusion of AGN. Again, we use both catalogues of physical quantities in this work. 

\subsection{SED fits of GAMA and DEVILS-D10}

Physical quantities of GAMA and DEVILS-D10 galaxies were estimated using SED fitting over all available bands in the extensive photometric catalogues. The fitting code of choice is \textsc{ProSpect} \citep{robothamProSpectGeneratingSpectral2020}, a novel, fully Bayesian, SED fitting software. Note that the details of \textsc{ProSpect} fits to the SEDs of GAMA and DEVILS galaxies is presented in \citet{bellstedt_prospect_2020} and \citet{thorneDeepExtragalacticVIsible2021a,thorneDeepExtragalacticVIsible2022a}, but we briefly summarise the process here. \textsc{ProSpect} was initially designed to generate realistic SEDs for synthetic galaxies in the \textsc{Shark} semi-analytic model \citep{lagos_shark_2018,lagos_far-ultraviolet_2019} given a non-parametric input star formation history (SFH). Due to its fully generative nature \textsc{ProSpect} can thus be used in fitting mode, minimising the residual between the input SED and the model SED for different permutations of the model parameters. At the simplest level, \textsc{ProSpect} relies on the conservation of energy to account for the life cycle of photons in galaxies \citep[e.g.,][]{conroyModelingPanchromaticSpectral2013}. 

The \citet{bruzual_stellar_2003} stellar spectral libraries are used in combination with the \citet{chabrier_galactic_2003} IMF. The observed stellar emission of galaxies is a product of the entire SFH of that galaxy; thus, \textsc{ProSpect} adopts parametric star formation histories that take the form of skewed-Normal distributions as a function of time, with a forced truncation of the SFH at $z \approx 11$ (look-back time $\approx 13$~Gyr) \citep{bellstedt_prospect_2020}. The redshift of this truncation was chosen to align with spectroscopic observations of GN-z11 that was, until recently, the oldest, spectroscopically confirmed galaxy to be observed \citep{oesch_remarkably_2016}. We note that recent observations with the NIRSpec instrument onboard the \textit{JWST} have placed GN-z11 at a slightly lower redshift than previously reported \citep[$z\approx 10.6$,][]{bunker_gnz11_2023}, and have also confirmed even older galaxies up to $z\approx 13$ \citep[e.g.,][]{curtis-lake_spectroscopy_2022}. As such, the truncation epoch could be regarded as a lower limit. 
In detail, the SFH is 

\begin{equation}
    \mathrm{SFR(t)_{snorm}} = \mathrm{\texttt{mSFR}} \times e^{\mathrm{\frac{-X(t)^{2}}{2}}},
    \label{eq:mass_func_snorm}
\end{equation}
where 
\begin{equation}
    \label{eq:mass_func_exp}
    \mathrm{X(t) = \left( \frac{t-\texttt{mpeak}}{\texttt{mperiod}}\right)(\mathit{e}^{\texttt{mskew}})^{asinh\left( \frac{t-\texttt{mpeak}}{\texttt{mperiod}} \right) }}.
\end{equation}
And with the additional truncation, the final function is 
\begin{multline}
    \mathrm{SFR(t)_{trunc} = SFR(t)_{snorm}} \times  \\ \mathrm{\left[ 1 - \frac{1}{2} \left[ 1 + erf\left( \frac{t-\mu}{\sigma \sqrt{2}}\right)\right]\right]},
    \label{eq:mass_func_snorm_trunc}
\end{multline}
where 

\begin{equation}
    \mathrm{\mu = \texttt{mpeak} + \frac{|\texttt{magemax}-\texttt{mpeak}|}{\texttt{mtrunc}}}
    \label{eq:mass_func_mu}
\end{equation}
and 
\begin{equation}
    \mathrm{\sigma = \frac{|\texttt{magemax}-\texttt{mpeak}|}{2 \times \texttt{mtrunc}}}.
    \label{eq:mass_func_sigma}
\end{equation}

In \Cref{eq:mass_func_mu,eq:mass_func_sigma}, $\texttt{magemax}$ is fixed at 13.4 Gyr ($z=11$) and $\texttt{mtrunc}$ is fixed at 2 Gyr. Therefore, this SFH uses four free parameters that describe the maximum SFR throughout the galaxy's life time (\texttt{mSFR}), the age at which that peak occurs (\texttt{mpeak}), the width of the underlying Normal distribution (\texttt{mperiod}) and the skewness of the Normal distribution (\texttt{mskew}). We refer to this function as \texttt{mass-func\_snorm\_trunc}, keeping with the conventions of \citet{robothamProSpectGeneratingSpectral2020,bellstedt_prospect_2020}.

Stellar light can then be attenuated by dust, and \textsc{ProSpect} implements this attenuation with the model of \citet{charlot_simple_2000} as
\begin{equation}
A(\lambda) = e^{-\tau (\lambda / 5500)^{-0.7}},
\label{eq:charlot_fall}
\end{equation}
where $\lambda$ has units of angstroms.
In this model, attenuation occurs in both the dense birth clouds of star formation and the surrounding interstellar medium (ISM). The two free parameters that control the attenuation are the optical depths of the birth clouds and the ISM. For stellar ages less than $10^{7}$ years old, the light is attenuated by both the birth cloud and the ISM whereas only the latter attenuates light for older stars. Attenuated light is then re-emitted as per the far infrared templates of \citet{dale_two-parameter_2014} where the two free parameters are $\alpha_{\mathrm{birth}}$ and $\alpha_{\mathrm{ISM}}$ that specify the slope of the power law of the radiation field that is heating the dust in the birth clouds and ISM respectively. In fitting the SEDs of galaxies, priors are imposed on the four free dust parameters to guide the fits to physical solutions as the dust parameters are treated as nuisance parameters \citep[see table 2 in ][for the priors on the dust parameters]{thorneDeepExtragalacticVIsible2021a}.

\citet{thorneDeepExtragalacticVIsible2022a} used the AGN model of \citet{fritz06agnmodel} incorporated into \textsc{ProSpect} to characterise the AGN component of GAMA and DEVILS galaxy SEDs. The \citet{fritz06agnmodel} model characterises the AGN as a combination of power laws from the UV-optical to the MIR. The MIR emission is the response of the dusty torus, which is described as a homogenised mixture of graphite and silicate particles. The dust geometry is assumed to be smooth (without clumps) and the grain size distributions are assumed to be power laws. There are seven 
parameters that determine the AGN properties in \textsc{ProSpect}, they are: (i) \texttt{AGNan} that controls the angle of observation, e.g., $\texttt{AGNan}=0^{\circ}$ is edge-on through the torus, (ii) \texttt{AGNlum} is the bolometric luminosity of the AGN, (iii) \texttt{AGNta} is the optical depth, (iv) \texttt{AGNrm} is the ratio of the outer to the inner torus radius, (v) \texttt{AGNbe} controls the radial distribution of the dust in the torus, (vi) \texttt{AGNal} controls the angular distribution of the dust in the torus, and (vii) \texttt{AGNct} is the opening angle of the torus. \texttt{AGNrm}, \texttt{AGNbe}, \texttt{AGNal} and \texttt{AGNct} are kept at fixed values in the fitting to mitigate for degeneracies between coupled AGN parameters; thus, the free parameters that were estimated from the fitting are \texttt{AGNan}, \texttt{AGNlum} and \texttt{AGNta} \citep[see Table 1 in ][for a full description of the AGN parameters]{thorneDeepExtragalacticVIsible2022a}.

It is also worth mentioning the metallicity prescription in \textsc{ProSpect} as variations in the implementation of metallicity can have significant impacts on reconstructed stellar masses and SFRs \citep{thorneDeepExtragalacticVIsible2021a,bellstedt_prospect_2020}. The metallicity evolution in \textsc{ProSpect} is linearly mapped to the stellar mass growth to self consistently track stellar evolution and chemical enrichment. As such, in modes of low star formation the metal enrichment rate is comparably low, but will increase in episodes of enhanced star formation.
The initial value at the genesis of the metallicity history is fixed to be 0.0001, which is the lowest metallicity template in the \citet{bruzual_stellar_2003} stellar libraries, and the final gas-phase metallicity at the epoch of observation, \texttt{Zfinal}, is the free parameter, which describes the metallicity of the gas from which the youngest stars formed.

In this work we use stellar mass, SFR and AGN bolometric luminosity estimates. Stellar masses are derived by integrating the SFH and subtracting the mass recycled in the ISM. Star formation rates are derived by taking the average of the SFH over the last $100$~Myr. The AGN bolometric luminosities are constrained by the AGN fitting.

The validity of \textsc{ProSpect} to estimate galaxy properties has been demonstrated through the excellent agreement between the literature and \textsc{ProSpect} derived stellar mass functions, main sequence of star formation and AGN bolometric light functions \citep[e.g.,][]{bellstedt_prospect_2020,thorneDeepExtragalacticVIsible2021a,thorneDeepExtragalacticVIsible2022a,thorne_devils_2022} and also against simulations where \textsc{ProSpect} has been shown to accurately recover simulated galaxy properties \citep[e.g.,][]{matias_prospect_2022,matias_prospect_2023}.

\begin{figure}
    \centering
    \includegraphics[width = \linewidth]{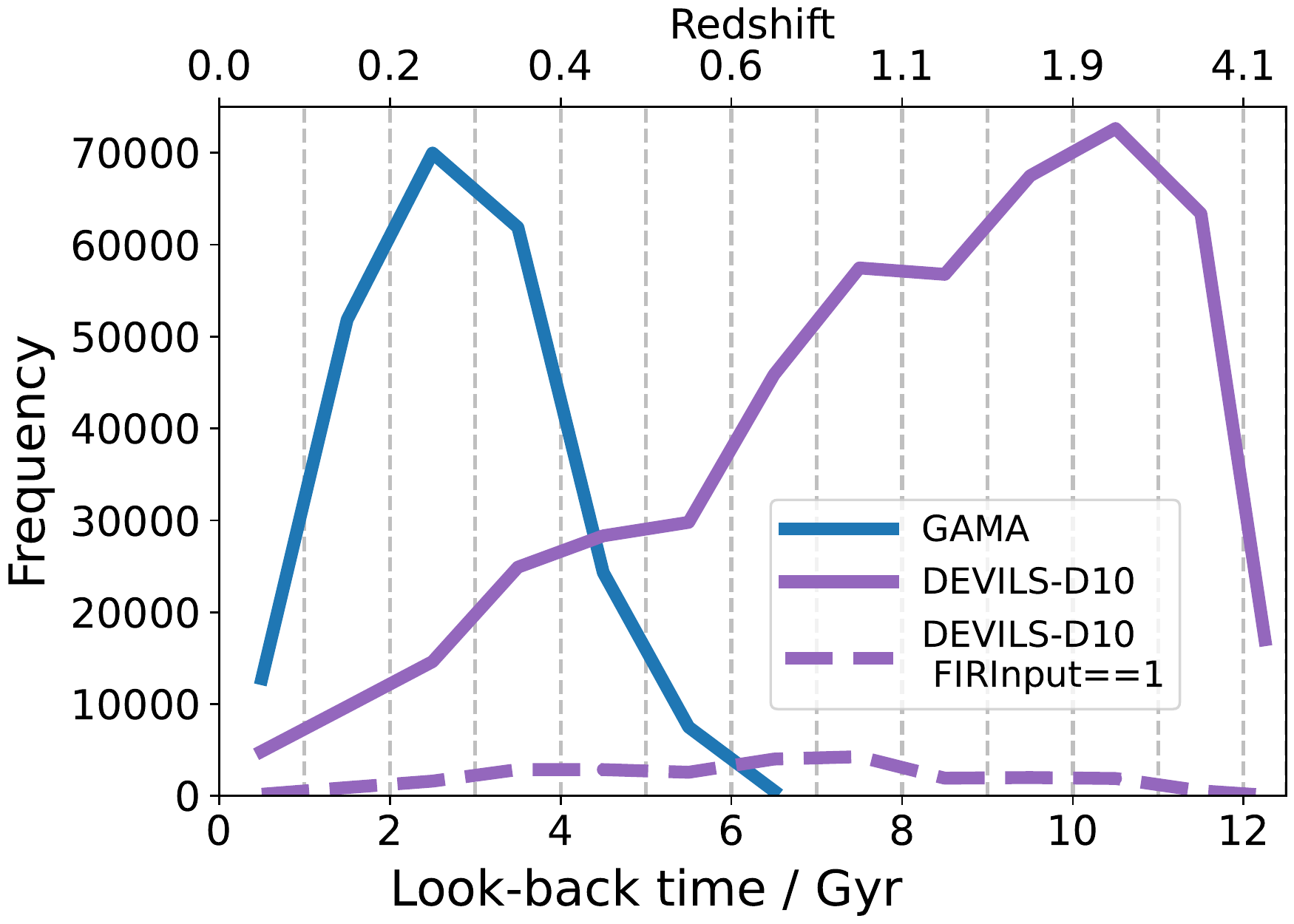}
    \caption{Frequency of galaxies per unit look-back time. Solid blue lines show the results for the GAMA sample and purple lines show the results for the DEVILS-D10 sample. The dashed purple line shows the frequency of DEVILS-D10 galaxies with FIR photometry. The dashed vertical lines show the 1 Gyr sized bin edges of look-back time. Note that the last bin however spans only 12 to 12.5 Gyr. }
    \label{fig:N-lbt}
\end{figure}

With our final sample of redshifts, stellar masses, SFRs and AGN bolometric luminosities for both GAMA and DEVILS-D10, we bin our sample into intervals of 1 Gyr in look-back time (our last bin spans look-back time $12.0 \to 12.5$~Gyr). \Cref{fig:N-lbt} shows the frequency of galaxies in GAMA and DEVILS-D10 in each of our 13 bins in look-back time. We purposefully truncate our GAMA sample to a limit of $z \lesssim 0.74$, which corresponds to a look-back time of $\approx 6$ Gyr, to suppress redshift uncertainties in the sample. We see that GAMA galaxies dominate the frequency of galaxies at look-back time $\lesssim 4.5$ Gyr, while DEVILS-D10 dominates thereafter. Thus, the benefit of using both DEVILS-D10 and GAMA is that we can access the full range of these distributions over $\approx 12.5$ Gyr of look-back time. 


\subsection{Additional AGN literature results}
\label{sect:agn_literature}
Because of cosmic variance, the area coverage of both GAMA and DEVILS-D10 is not wide enough to sample the brightest AGN in the Universe meaning that the AGN luminosity distribution is unconstrainable with these two data sets alone. Thus, we augment our GAMA and DEVILS-D10 data with the AGN bolometric luminosity functions of \citet{shen_bolometric_2020}. These are compiled from a variety of multi-wavelength quasar luminosity functions, covering the hard X-Rays ($\approx 1.5 \to 6$ \SI{}{\angstrom}), soft X-rays ($\approx 6 \to 25$ \SI{}{\angstrom}), UV ($\approx \SI{1450}{\angstrom}$), B band  ($\approx \SI{4400}{\angstrom}$) and mid-IR ($\approx 8\times 10^{4} \to 15 \times 10^{4}$ \SI{}{\angstrom}). \citet{shen_bolometric_2020} also apply dust extinction corrections and bolometric conversions, meaning that the inclusion of this dataset is consistent with our results from GAMA/DEVILS-D10.

We use their `global fit A' luminosity functions that are double-power law fits to the distributions and where the parameters of the double-power law depend on redshift. The first step to fold in these data, then, is to compute the values of the parameters of the double-power law at the median redshift in each of the GAMA/DEVILS-D10 redshift bins. Next, we compute the AGN bolometric luminosity distribution on a grid of luminosities of $\log_{10}(\mathrm{L_{bol}}/\mathrm{erg \, s^{-1}}) = 35 \to 50$ in steps of 0.5. The final step is to concatenate these additional distributions with our own. We calculate the maximum non-zero luminosity bin for our GAMA/DEVILS-D10 results, shift that limit by $2$~dex fainter and add the \citet{shen_bolometric_2020} distributions beyond this point toward the bright end.

The result of this exercise is that the constraining power for the faint end of the AGN bolometric luminosity distribution function comes from the GAMA/DEVILS-D10 results, while it comes from the \citet{shen_bolometric_2020} distributions at the bright end. The benefit of the GAMA/DEVILS-D10 data at the faint end is motivated by the potential for even the deepest X-ray surveys to miss a significant population of obscured AGN. The faint end of the AGN bolometric luminosity distributions that \cite{shen_bolometric_2020} use in their work is mostly dominated from the contributions of hard X-rays at $\mathrm{L_{bol}} \lesssim 10^{43} \, \mathrm{erg \, s^{-1}}$. \citet{thorne_devils_2022} performed a positional match between \textsc{ProSpect} identified AGN and \textit{Chandra} X-ray sources in the COSMOS field from \citet{marchesi_2016_agn_cosmos}, finding that there are $\sim 8000$ \textsc{ProSpect} identified AGN with X-ray counterparts above the sensitivity threshold of \textit{Chandra} that were not matched with known X-ray sources. This suggests that either \textsc{ProSpect} is overestimating their bolometric luminosities or there are potentially Compton-thick, obscured AGN that will be undetectable in deep \textit{Chandra} surveys.




\section{Methods}
\label{sect:methods}

\subsection{SFR and AGN replacement}
\label{subsect:replace}
\citet{thorneDeepExtragalacticVIsible2022a} found that the far infrared input for galaxy SED fitting with \textsc{ProSpect} is critical for determining accurate SFRs and AGN bolometric luminosities. The GAMA photometric data set contains far infrared (FIR) photometry input for all galaxies, whereas the DEVILS-D10 data set only includes FIR measurements for galaxies that are brighter than 21.2 mag in the Y band. Without FIR measurements the AGN component is unconstrained, which can lead to catastrophic underestimation of the SFR, especially for the galaxies that \textsc{ProSpect} estimates to have a $\gtrsim 10$ per cent fraction of the SED dedicated to AGN flux. To ensure that we are not biased by unconstrained AGN outputs we replace the stellar masses and SFRs of the galaxies that are missing FIR photometry with the stellar masses and SFRs derived with \textsc{ProSpect} without AGN templates \citep{thorneDeepExtragalacticVIsible2021a}. 
For the AGN bolometric luminosities of galaxies missing FIR photometry, we do not have a complementary data set with which we can replace them, as was the case with their SFRs. Instead, we experiment with setting their AGN bolometric luminosities to either 0 AGN luminosity or the lower bound of the \textsc{ProSpect} derived AGN luminosity. Replacing the AGN luminosities with 0 represents the minimum AGN solution for the galaxies missing FIR photometry. As this replacement does not allow for any of these galaxies to retain any AGN flux, replacing their luminosities with the \textsc{ProSpect} lower bound AGN allows us to present a more physically meaningful AGN solution for these same galaxies. In essence, the motivation behind these replacements is to encompass a range of viable AGN solutions (we return to the effect of these replacements on the cosmic AGN bolometric luminosity density in \Cref{subsubsect:cagn}). We reiterate that this replacement is only applicable to the DEVILS-D10 data set. 

\begin{table}
\begin{center}
\begin{tabular}{|p{0.35\linewidth} | p{0.55\linewidth}|}
\hline
   Replacement name  & Description \\
   \hline
   \hline
    SFR$_{\mathrm{Pro-Stellar}}$ & SFRs determined from the pure stellar component run of \textsc{ProSpect}, i.e., no AGN component was included to model the galaxy SEDs \citep[e.g.,][]{thorneDeepExtragalacticVIsible2021a}. \\
    \hline
    SFR$_{\mathrm{Pro-Stellar+AGN}}$& SFRs of galaxies determined from the \textsc{ProSpect} run that included both AGN and stellar components \citep[e.g.,][]{thorneDeepExtragalacticVIsible2022a}. \\
    \hline
    SFR$_{\mathrm{Hybrid}}$ & For galaxies missing FIR photometry we replace the Pro-Stellar+AGN SFRs with the Pro-Stellar SFRs.\\
    \hline
    AGN$_{\mathrm{Pro}}$ & AGN bolometric luminosities straight out of the \textsc{ProSpect} fits to the GAMA and DEVILS-D10 galaxy SEDs. \\
    \hline
    AGN$_{\mathrm{LB}}$ & AGN bolometric luminosities where for GAMA and DEVILS-D10 galaxies missing FIR photometry we replace the AGN$_{\mathrm{Pro}}$ AGN bolometric luminosities with the lower bound AGN bolometric luminosity from \textsc{ProSpect}. \\
    \hline
    AGN$_{\mathrm{0}}$ & AGN bolometric luminosities where for GAMA and DEVILS-D10 galaxies missing FIR photometry we replace the AGN$_{\mathrm{Pro}}$ bolometric luminosities with 0 AGN bolometric luminosity. \\
    \hline
\end{tabular}
\end{center}
\caption{Description of our various replacements to either the SFRs or AGN bolometric luminosities. The AGN subsets are chosen to encompass a range of viable AGN bolometric luminosity solutions for the galaxies missing FIR photometry.}
\label{tab:replace_defn}
\end{table}

For clarity, we show in \Cref{tab:replace_defn} a tabular description of all the different SFRs that we use in this work depending on what variant of \textsc{ProSpect} was used to fit the galaxy SEDs and determine SFR. We also describe the different AGN bolometric luminosities that we use depending on whether we replace the \textsc{ProSpect} determined AGN bolometric luminosity with the lower bound or 0. We refer to these descriptions throughout the text.

\subsection{Completeness selections and concatenation}
The extension towards the smallest stellar masses, SFRs, or faintest AGN bolometric luminosities is not possible with sensitivity limited surveys because of the Malmquist bias \citep[e.g.,][]{weigelStellarMassFunctions2016}. If we do not account for this incompleteness then we would be underestimating the cosmic density of stellar mass, SFR and AGN luminosity distributions. 

The stellar mass distribution is well described as monotonically increasing toward low stellar masses beyond a characteristic ``knee'' \citep[e.g.,][]{baldry_mf_2012,taylor_mf_2015}. As such we can pin any turn over of the stellar mass distribution at low stellar mass to incompleteness. We note that there are theoretical reasons as to why the stellar mass distribution rises monotonically toward low stellar mass \citep[e.g.,][]{press_schechter_1974} but the same is not necessarily true for SFR and AGN bolometric luminosity. Nevertheless, we apply this thinking to not only stellar masses, but SFRs and AGN luminosities as well, motivated by their established tight, correlations with stellar mass \citep[e.g.,][]{noeskeStarFormationAEGIS2007,salimUVStarFormation2007, faberVelocityDispersionsMasstolight1976, ferrareseFundamentalRelationSupermassive2000}. 

Following \citet{driverGAMAG10COSMOS3DHST2018}, we calculate the stellar mass distributions of both the DEVILS-D10 and GAMA data sets in the range $\mathrm{6.5 \leq \log_{10}(M_{\star}/M_{\odot}) \leq 12.5}$ and in bins of 0.2 dex. We then find the peak of the distributions and define this peak as the completeness limit. We repeat this same process for the SFRs in the range $\mathrm{-7.5 \leq \log_{10}(SFR/M_{\odot}yr^{-1}) \leq 7.5}$ and the AGN bolometric luminosities in the range $\mathrm{30 \leq \log_{10}(AGN lum/erg \, s^{-1}) \leq 60}$; we use the same bin size of 0.2 dex for both. \footnote{We use broad limits of SFR and AGN bolometric luminosity only for the sake of completeness. The maximum SFR$/\mathrm{M_{\odot}yr^{-1}}$ (AGN bolometric luminosity/$\mathrm{erg \, s^{-1}}$) in our joint GAMA/DEVILS-D10 sample is $\approx 10^{4}$ ($\approx 10^{48.5}$) and the minimum is $\approx 0$ ($10^{35}$).} 

\begin{figure}
    \centering
    \includegraphics[width = \linewidth]{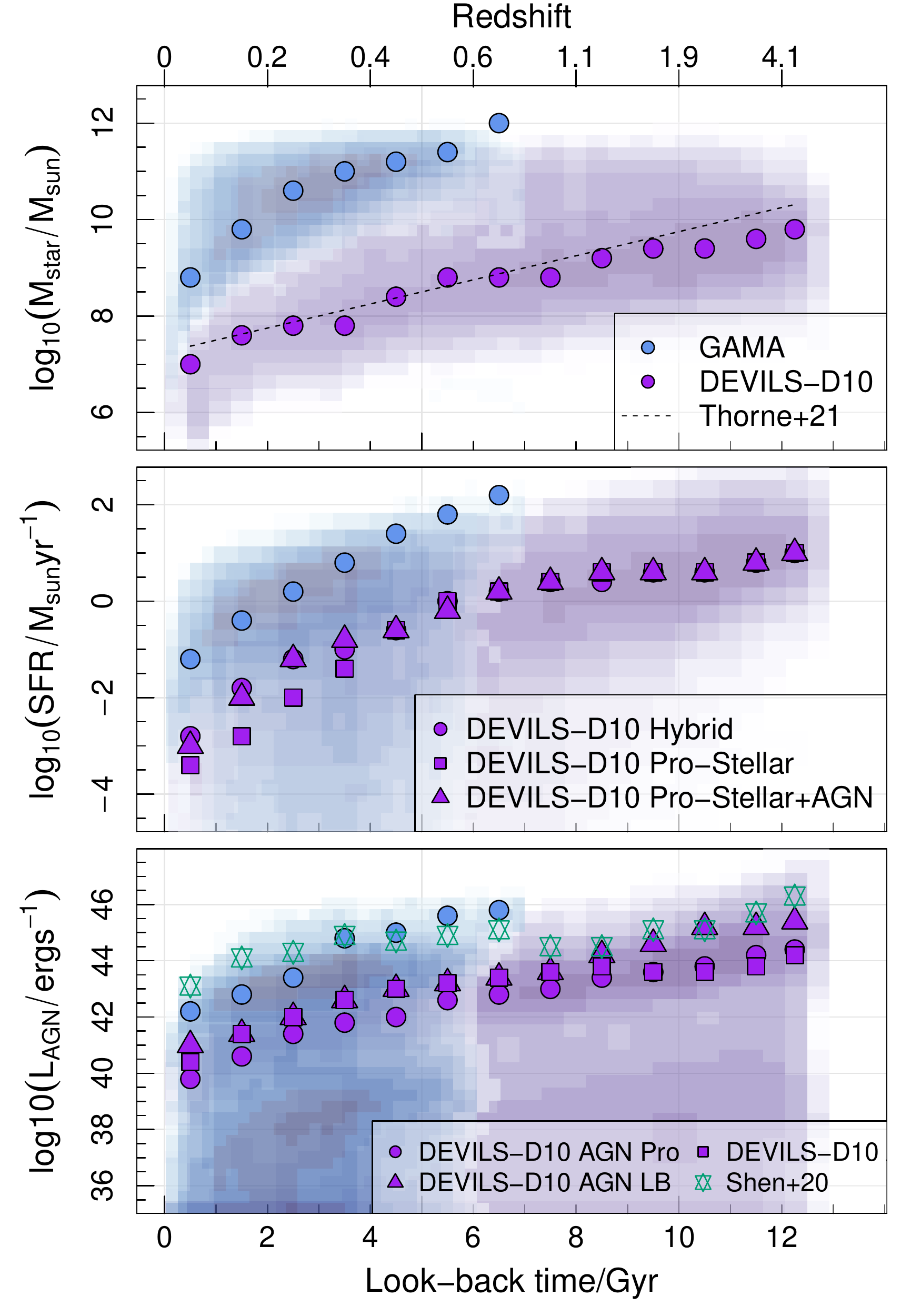}
    \caption{Completeness limits per unit look-back time for stellar mass (top), SFR (middle) and AGN bolometric luminosity (bottom). Completeness is defined to be the point at which the number density distributions turn over. Blue points refer to the GAMA sample and purple points refer to the DEVILS-D10 sample. The shades are 2D frequency distributions. We show three different completeness limits each for the DEVILS-D10 SFRs and AGN bolometric luminosities according to the replacement logic that we outline in \Cref{subsect:replace} and \Cref{tab:replace_defn}. As all GAMA galaxies have FIR input, we do not need to replace the physical quantities between \textsc{ProSpect} runs. The green six-pointed star symbols show the limits that we use for the AGN bolometric luminosity distributions from \citep{shen_bolometric_2020}. The dashed line in the top panel is the stellar mass completeness definition used in \citet{thorneDeepExtragalacticVIsible2021a}.}
    \label{fig:completeness_limits}
\end{figure}

\Cref{fig:completeness_limits} shows the evolution of the stellar mass, SFR and AGN luminosity completeness limits used for GAMA and DEVILS-D10. We draw attention to our agreement with the stellar mass completeness used by \citet{thorneDeepExtragalacticVIsible2021a} for the DEVILS-D10 sample. 



\subsection{Large scale structure correction}
\label{subsect:LSS}
Cosmic variance can bias constraints of the CSFH and CAGNH. For example, a dip in the cosmic stellar mass density at $z\approx0.5$ was reported for galaxies observed in the COSMOS field \citep{driverGAMAG10COSMOS3DHST2018}, the cause of which was pinned down to an underdensity in the field at that redshift. DEVILS-D10 is also in COSMOS and therefore is subject to this large scale structure effect. We correct for this by calculating the cosmic stellar mass history for the combined GAMA and DEVILS-D10 set. We assume that the stellar mass density monotonically and smoothly increases as a function of decreasing redshift, and so we pin down any deviation from this shape to cosmic variance. 

We calculate the cosmic stellar mass density as 
\begin{equation}
    \rho_{\mathrm{M_{\star}}} = \int_{0}^{\infty}\phi(\mathrm{M_{\star}})\times \mathrm{M_{\star} d M_{\star}}
    \label{eq:csmd}
\end{equation}
where $\phi(\mathrm{M_{\star}})$ is the stellar mass function. We derive an empirical stellar mass function by fitting the stellar mass distribution with a seventh order smooth spline, which was found to give the best fit according to the reduced chi-squared statistic when experimenting with orders $4\to9$. We weight each data point by the inverse of the variance, $\sigma^{2}$. We use the Poisson errors and a 20 per cent error floor that we add in quadrature to account for systematic uncertainties and cosmic variance. We also follow the approach of \citet{driverGAMAG10COSMOS3DHST2018} and use the information of null data in high stellar mass bins to ensure that our integrals are convergent. Specifically, the first stellar mass bin with 0 elements is set to be 5 per cent of the minimum of the stellar mass distribution, which ensures that the resulting stellar mass density converges and is not biased by the extrapolation of the spline fits. We use a Monte-Carlo approach to estimate an uncertainty by perturbing the data points about their errors and refitting the distribution 101 times. We then integrate these fits to find the median and 16-84 percentile range of the stellar mass density. We integrate from $\mathrm{0 \leq \log_{10}(M_{\star}/M_{\odot}) \leq 100}$. 

\begin{figure}
    \centering
    \includegraphics[width = \linewidth]{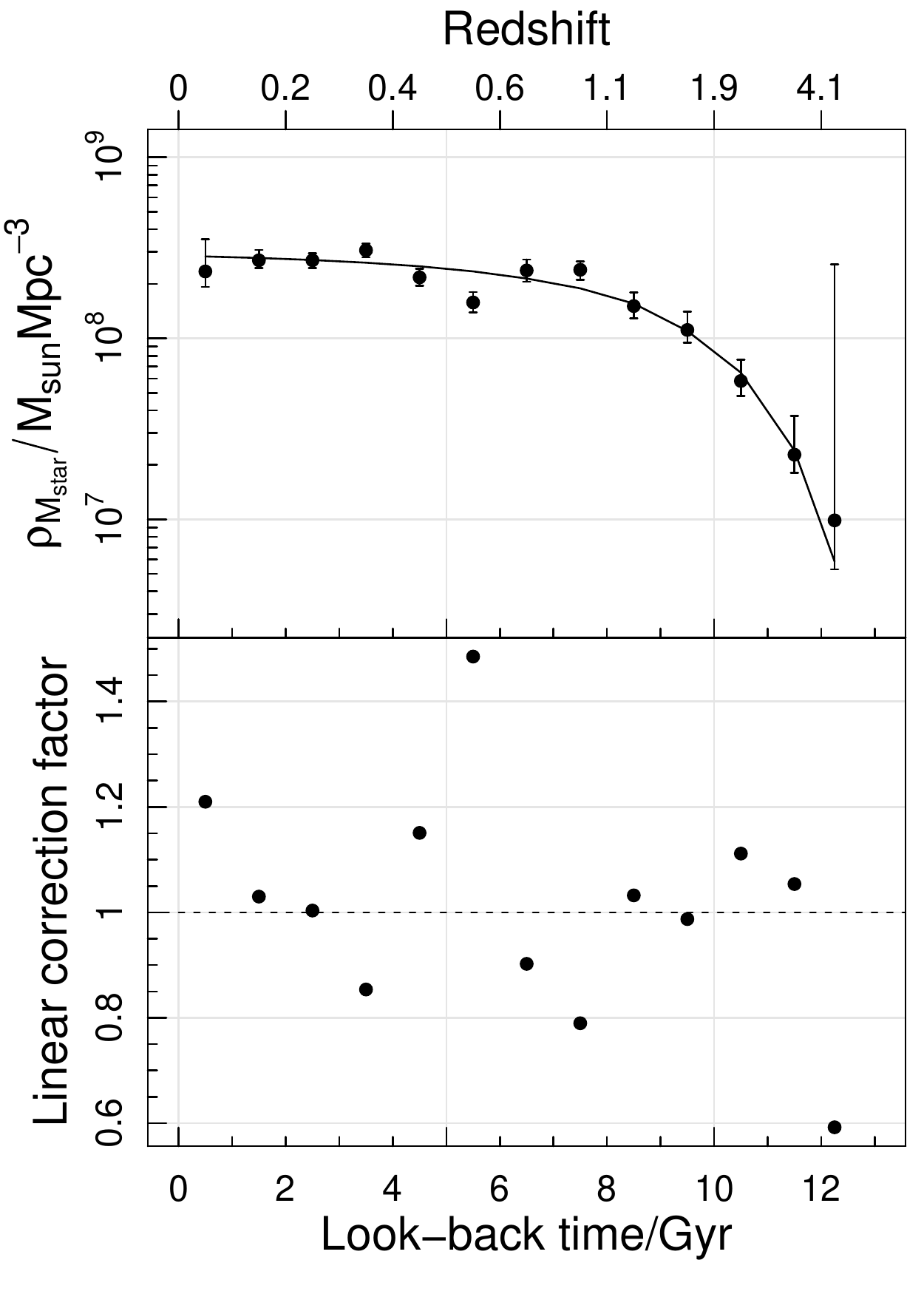}
    \caption{\textit{Top: } Cosmic stellar mass history. The solid lines shows a fit to the data points where we fit only to the normalisation constant, $\psi_{\mathrm{norm}}$, in \Cref{eq:md14_csmd}. \textit{Bottom: } Linear correction factor obtained by dividing the value of the fit by the value of the data point as a function of lookback time. The dashed line is the unitary line where the correction factor has no effect on the data point.}
    \label{fig:LSS}
\end{figure}
The top panel of \Cref{fig:LSS} shows the resulting cosmic stellar mass history. The error bars on the points are derived from the 16-84 percentiles of the Monte-Carlo fits. A local minimum is clearly seen around at look-back time $\approx 5$~Gyr that is evidence for the underdensity in the COSMOS field. To account for this, we fit a smooth function and define any offset from the fit (correction = fit/data) as the cosmic variance correction. We use this correction throughout in our analysis of the CSFH and CAGNH.

\begin{multline}
    \rho_{M_{\star}}(z) = \\ (1-R)\int_{z}^{\infty} \left[ \psi_{\mathrm{norm}}\frac{(1+z')^{2.7}}{1 + [(1+z')/2.9]^{5.6}} \right] \frac{\mathrm{d}z'}{H(z')(1+z')},
    \label{eq:md14_csmd}
\end{multline}

\noindent Eq. \ref{eq:md14_csmd} shows the smooth function that we use to fit the cosmic stellar mass history, where $R$ is the return fraction, $\mathrm{\psi_{norm}}$ is the normalisation and $H(z')$ is the redshift dependent Hubble parameter. The form of the function inside the square brackets in \Cref{eq:md14_csmd} was determined by \citet{madauCosmicStarFormationHistory2014}. We adopt their fit parameters but leave the normalisation as a free parameter, i.e., we fit for $\mathrm{\psi_{norm}}$. We also assume a return fraction $R=0.5$, which is the fraction of mass that is put back into the ISM for each episode of star formation. $R$ will depend on the choice of IMF, but for the purposes of fitting only for the normalisation of the cosmic stellar mass history its exact numerical value is unimportant. The result of this fit is shown by the line in the top panel of \Cref{fig:LSS}, and the bottom panel shows the final correction factor that we use at each look-back time bin. 

\begin{figure}
    \centering
    \includegraphics[width = \linewidth]{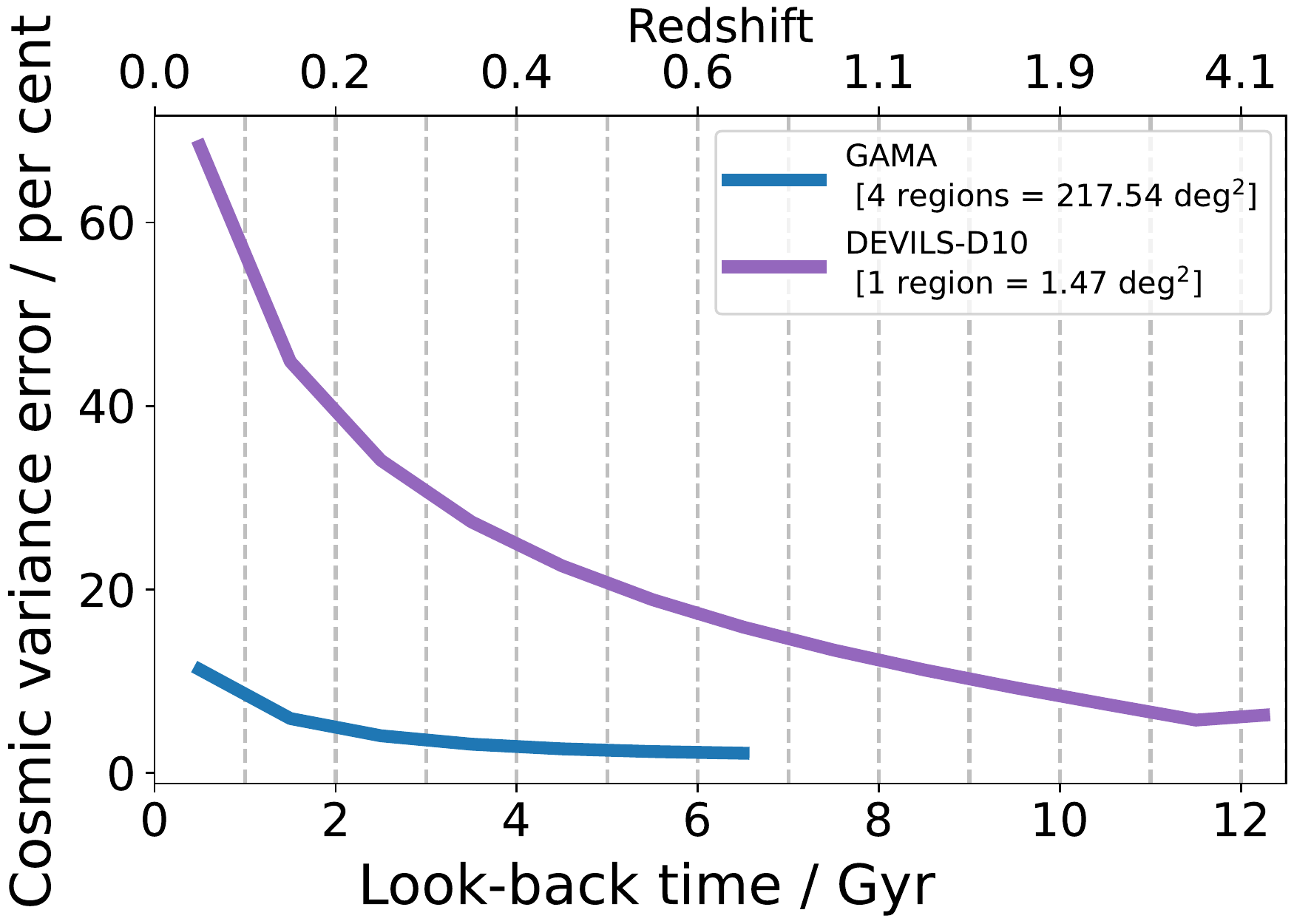}
    \caption{Cosmic variance error as per the prescription of \citet{driver_quantifying_2010} in each of our bins of look-back time. Results for the DEVILS-D10 sub-survey are shown in purple and the results for the GAMA survey are shown in blue.}
    \label{fig:cosvar}
\end{figure}

We note that the 20 per cent error floor plays an important role in our spline fitting. \Cref{fig:cosvar} shows the approximate per cent error caused by cosmic variance for GAMA and DEVILS-D10, which we estimate using the prescription of \citet{driver_quantifying_2010}. The error in GAMA is $\lesssim 10$ per cent in all of our look-back time bins because of its large area of $\approx 217.54 \, \mathrm{deg^{2}}$ distributed over four independent sight lines. The error associated with DEVILS-D10 is far worse with a maximum error at look-back time $\approx0$~Gyr as high as $\approx70$ per cent due to the narrower area of $\approx 1.47 \, \mathrm{deg^{2}}$ distributed over only a single sight line. As  demonstrated in \Cref{fig:N-lbt} however, GAMA dominates the number counts at look-back time $\lesssim 4.5$ Gyr. From then on, the cosmic variance error associated with DEVILS-D10 is $\lesssim 20$ per cent. As such, we use this value of 20 per cent as an error floor to encompass cosmic variance, which is not implicit in the Poisson error. 

\subsection{Spline fitting SFR and AGN distributions}
\begin{figure*}
    \centering
    \includegraphics[width = 0.99\textwidth]{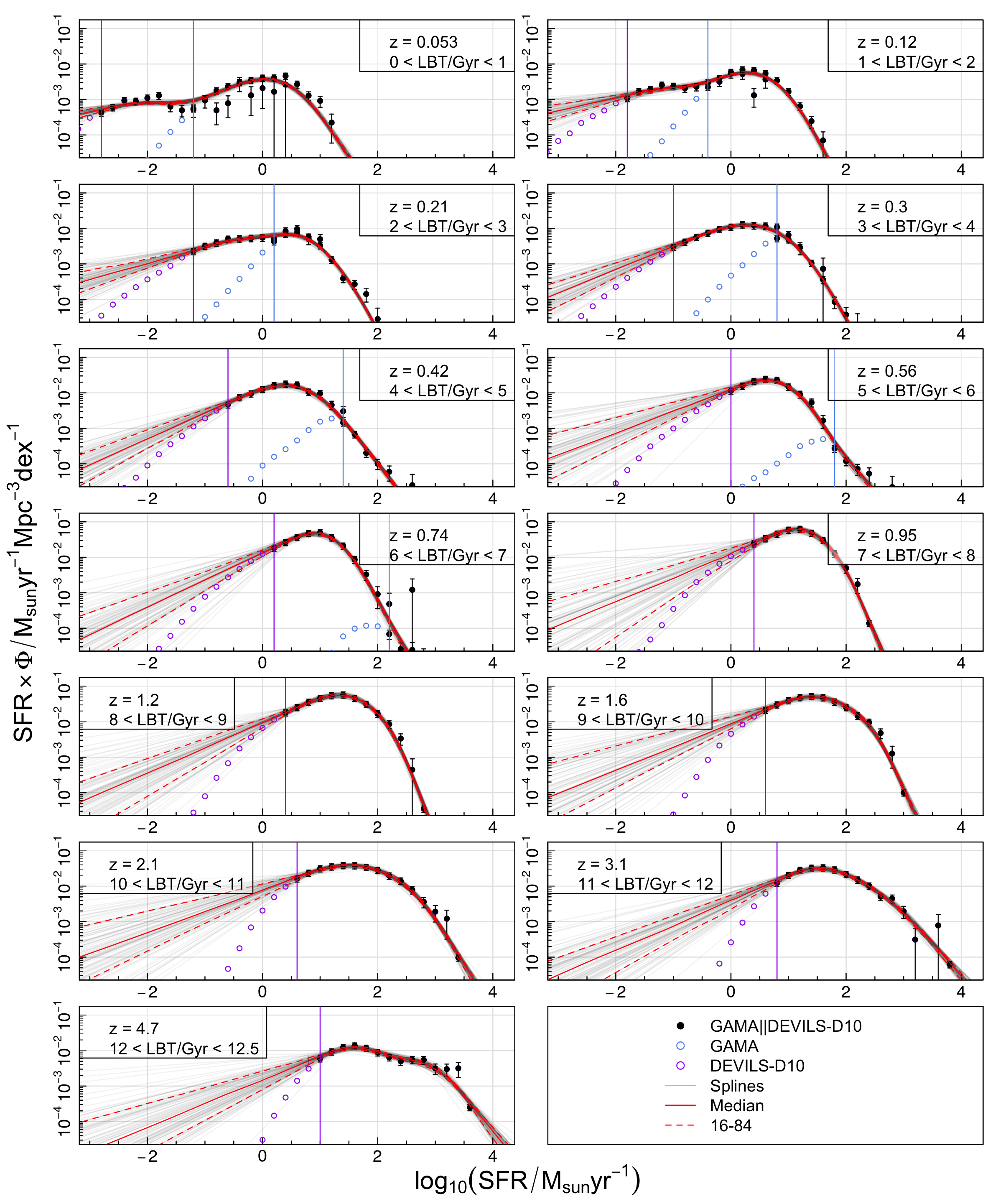}
    \caption{SFR density distributions from look-back time $0 \to 12.5$~Gyr ($z=0\to5$) as labelled for GAMA (blue points) and DEVILS-D10 (purple points) for the $\mathrm{SFR_{Hybrid}}$ sample, as per the definitions in \Cref{subsect:replace} and \Cref{tab:replace_defn}. Coloured, vertical lines mark the completeness limits of the distributions. We only use GAMA up to a limit of look-back time $\approx 6\to7$~Gyr ($z\approx 0.74$). Faint grey lines show 101 Monte-Carlo smooth spline fits to the distributions where we have perturbed the points about their normal errors. Red, solid lines show the median of these spline fits and the red, dashed lines show the 16-84 percentiles. The area under these curves gives the CSFRD. To mitigate incompleteness we only fit the data points in either GAMA and DEVILS-D10 that are to the right of the vertical lines, and we mark those points with black dots and error bars. Included in the error bar is the Poisson error and a 20 per cent error floor added in quadrature. In each panel, the redshift in the legend is the median redshift in the look-back time bin.}
    \label{fig:sfr_density}
\end{figure*}
\begin{figure*}
    \centering
    \includegraphics[width = 0.99\textwidth]{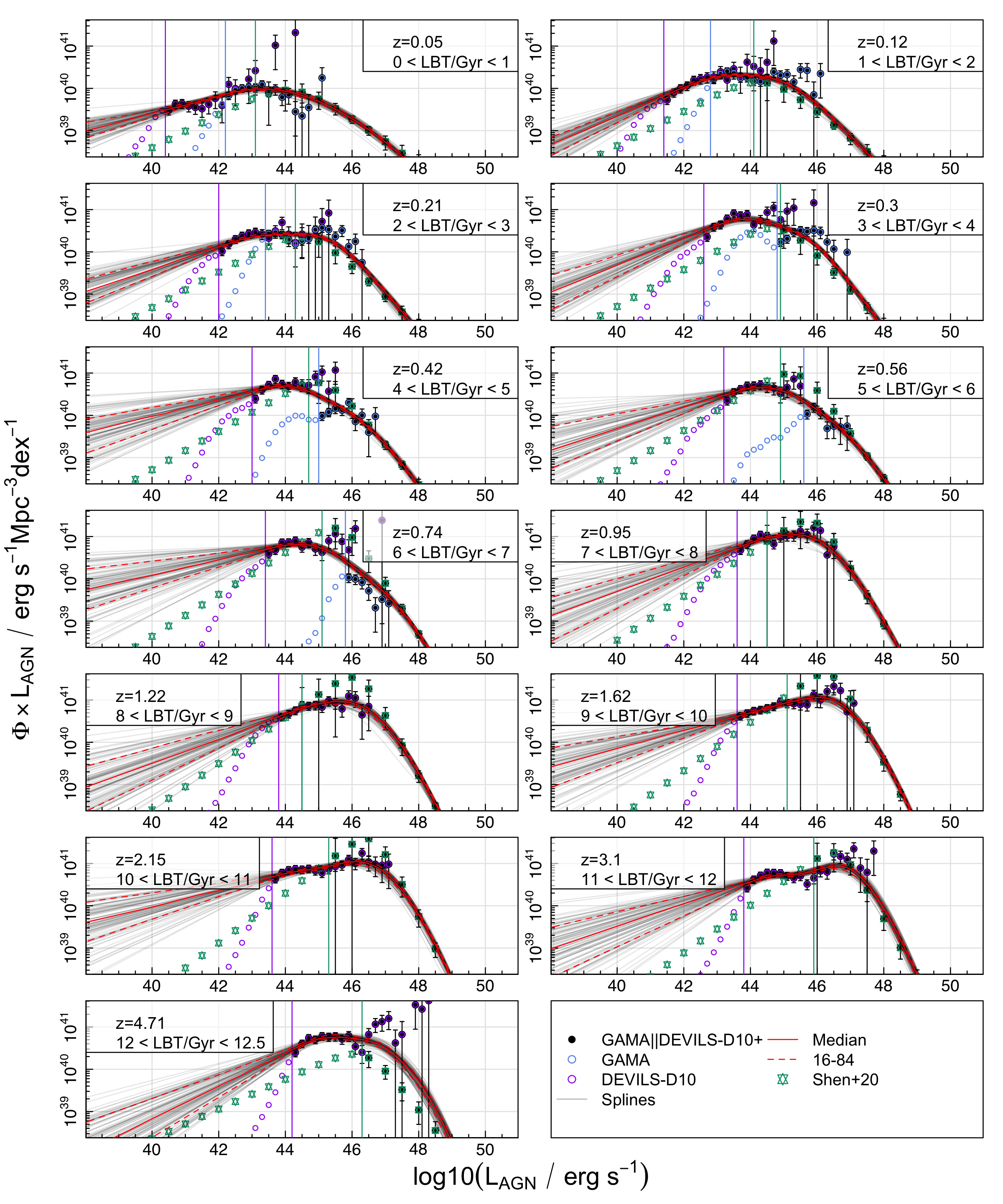}
    \caption{AGN luminosity density distributions from look-back time $0 \to 12.5$~Gyr ($z=0\to5$) as labelled for GAMA (blue points) and DEVILS-D10 (purple points) for the AGN$_{\mathrm{LB}}$ sample, as per the definitions in \Cref{subsect:replace} and \Cref{tab:replace_defn}. We only use GAMA up to a limit of look-back time $\approx 6\to7$~Gyr ($z\approx 0.74$). We also include the light functions of \citet{shen_bolometric_2020}, which we show with green stars. Coloured, vertical lines mark the completeness limits of the distributions. Faint grey lines show 101 Monte-Carlo smooth spline fits to the distributions where we have perturbed the points about their normal errors. Red, solid lines show the median of these spline fits and the red, dashed lines show the 16-84 percentiles. The area under these curves gives the cosmic AGN bolometric luminosity density. To mitigate incompleteness we only fit the data points that are to the right of the vertical lines, and we mark those points with black crosses and error bars. Included in the error bar is the Poisson error and a 20 per cent error floor added in quadrature. In each panel, the redshift in the legend is the median redshift in the look-back time bin.}
    \label{fig:agn_density}
\end{figure*}
To estimate the cosmic SFR and AGN bolometric luminosity density we use a similar prescription as presented in \Cref{eq:csmd}. We derive the SFR and bolometric AGN luminosity functions first, and then fit the SFR and AGN luminosity distributions with fifth order smooth splines, weighting again by the inverse of the variance added in quadrature to a 20 per cent error floor. Fifth order splines were found to give the best fits to the data, meaning that the extra degrees of freedom that we used to fit the stellar mass distributions were not necessary when fitting the SFR and AGN bolometric luminosity distributions. We also place a high significance, low-value data point anchor in a high SFR/AGN luminosity bin. Again, we constrain an error bound by performing a Monte-Carlo experiment: fitting the distributions 101 times after perturbing the points about their uncertainties. 

\Cref{fig:sfr_density} shows all 13 of our SFR density distributions for the joint $\mathrm{SFR_{Hybrid}}$. The spline fits are forced to be linear when extrapolating beyond the data meaning that these fitted curves are indeed strictly decreasing beyond the peaks of the distributions and the resulting integrals will be convergent. We also show the completeness limits for each of the two data sets as vertical lines. The spline fits experience greater variance to the left of these limits at the faintest end because the precise shape of the distribution function is unknown. This allows us to constrain an error bound on the integrated CSFH that encompasses a range of possible shapes to the SFR distribution function. The curves for the $\mathrm{SFR_{Pro-Stellar}}$ and $\mathrm{SFR_{Pro-Stellar+AGN}}$ show similar bounded shapes as \Cref{fig:sfr_density}.

\Cref{fig:agn_density} show all 13 of our AGN bolometric luminosity density distributions for the $\mathrm{AGN_{LB}}$ sample combined with the light functions of \citet{shen_bolometric_2020}. The benefit of the inclusion of this additional data set, which we discussed in \Cref{sect:agn_literature}, is apparent in most of our look-back time bins, as the bright end of the distribution is bound and the area under the curves are convergent, which is not the case when only using GAMA and DEVILS-D10. Much like the SFR density distributions, the variance of the splines is greater at the faint end as the precise shape of the AGN bolometric luminosity function there is unknown. Again, the curves for the $\mathrm{AGN_{Pro}}$ and $\mathrm{AGN_{0}}$ samples are similar to \Cref{fig:agn_density}. 

We then take the median and 16-84 percentiles of the Monte-Carlo spline fits and integrate the extrapolations to determine cosmic densities and an uncertainty quantile. 
We integrate from $\mathrm{-50 \leq \log_{10}(SFR/M_{\odot}yr^{-1}) \leq 50}$ and $\mathrm{0 \leq \log_{10}(AGN lum/erg \, s^{-1}) \leq 100}$ for the SFR and AGN bolometric luminosities respectively to encompass a wide domain of SFRs and AGN bolometric luminosities.

\section{Results and discussion}
\label{sect:results-and-discussion}

\subsection{Impact of AGN in \textsc{ProSpect}}
\label{subsect:impact_agn}
\begin{figure*}
    \centering
    \includegraphics[width = \textwidth]{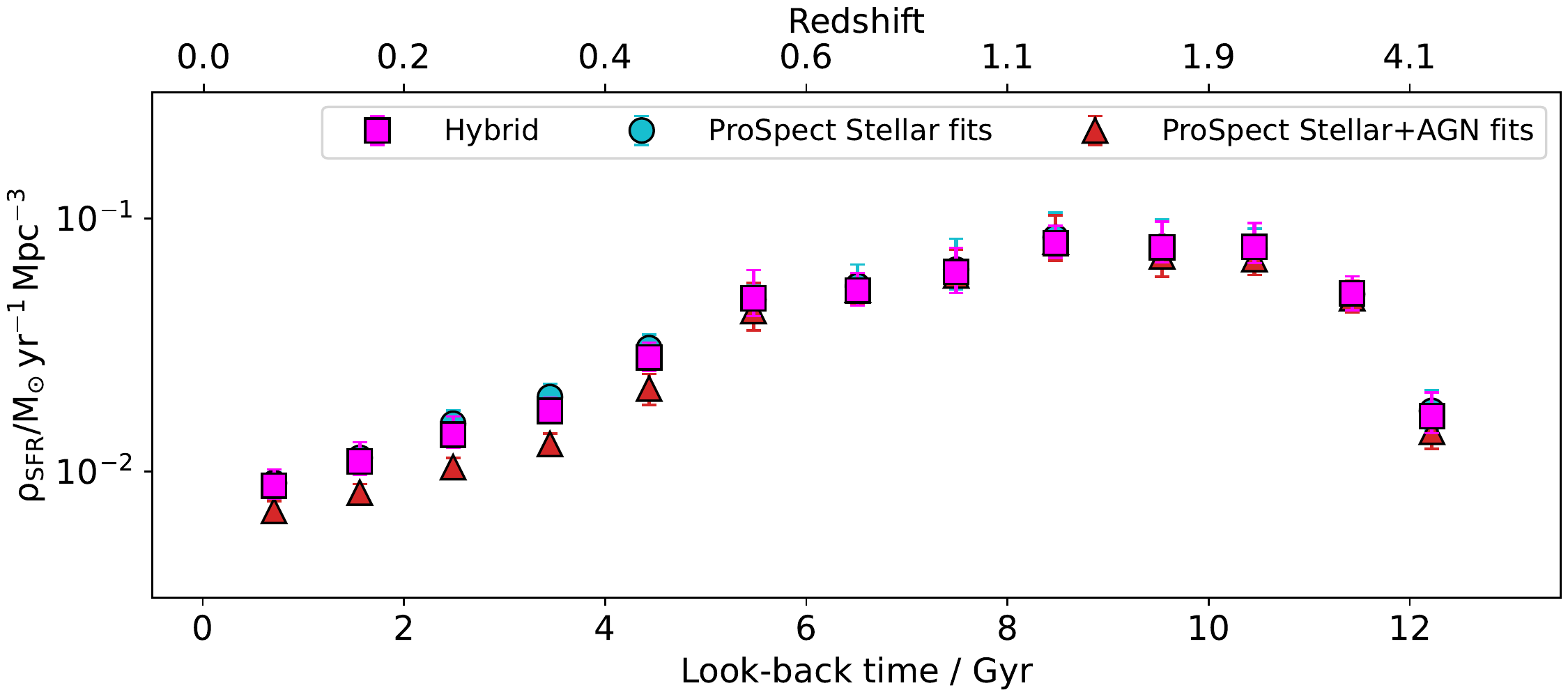}
    \caption{CSFH for the combined set of GAMA and DEVILS-D10. Blue points show the result for the $\mathrm{SFR_{Pro-Stellar}}$ sample. Red points show the result for the $\mathrm{SFR_{Pro-Stellar+AGN}}$ sample. Magneta points show the results for the $\mathrm{SFR_{Hybrid}}$ sample.
    }
    \label{fig:agn_no_agn}
\end{figure*}

\Cref{fig:agn_no_agn} shows the CSFH for our three possible configurations of GAMA and DEVILS-D10 SFR as per the definitions in \Cref{subsect:replace} and \Cref{tab:replace_defn}. We have applied the LSS correction that has been derived from the hybrid sample. We find very little variation between the stellar mass distributions for each of the three different replacement subsets of stellar mass, which means that the LSS correction will thus be similar between them. 

The CSFH resulting from the SFR$_{\mathrm{Pro-Stellar+AGN}}$ sample is lower by $\approx 0.2$ dex at look-back time $\lesssim 6$ Gyr compared to the SFR$_{\mathrm{Pro-Stellar}}$ sample. The magenta points show our hybrid sample where for the galaxies that are missing FIR photometry we replace their SFRs from the SFR$_{\mathrm{Pro-Stellar+AGN}}$ sample with the SFRs from the SFR$_{\mathrm{Pro-Stellar}}$ sample. The 0.2 dex deficit is thus a result of the AGN component of the SED being unconstrained for galaxies that lack FIR input that erroneously diverts flux away from star formation toward AGN \citep{thorneDeepExtragalacticVIsible2022a}. 

At look-back time $\gtrsim 7$~Gyr ($z>1$) the number of objects with registered FIR photometry is far less than the rest of the sample, and so the AGN component of the SED is more unconstrained than at lower redshifts. Despite this, the results from the SFR$_{\mathrm{Pro-Stellar+AGN}}$ sample are in agreement with the results from the SFR$_{\mathrm{Pro-Stellar}}$ sample for look-back time $> 6$ Gyr. 

\begin{figure}
    \centering
    \includegraphics[width = \linewidth]{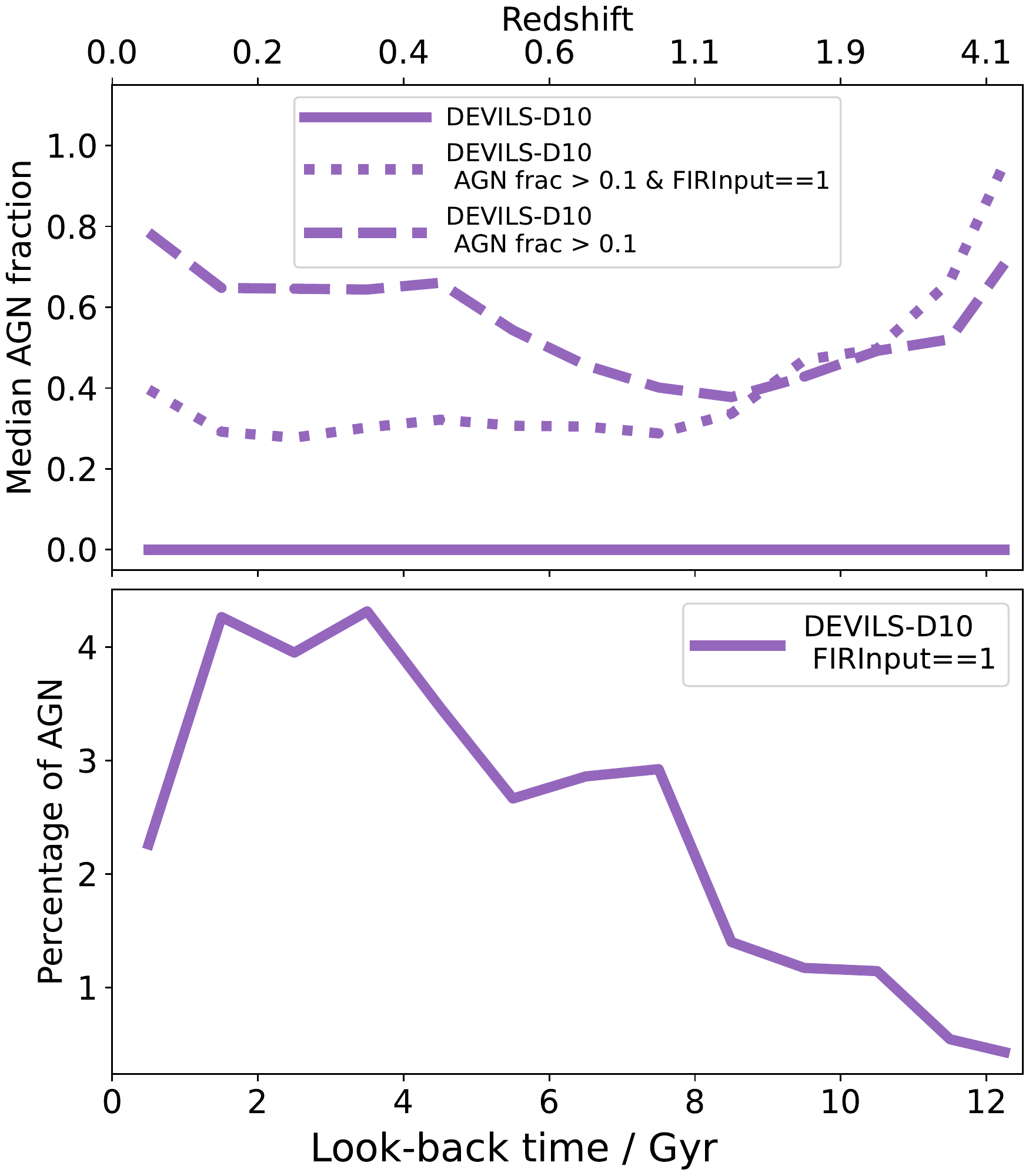}
    \caption{\textit{Top:} Median AGN fraction per unit look-back time for the DEVILS-D10 data set. The dashed line shows the result only for galaxies with significant AGN, i.e., AGN fraction > 0.1, where AGN fraction is defined as the flux contribution between 5 and 20 $\mu$m over the total SED, as per the definition in \citet{thorneDeepExtragalacticVIsible2022a}. The dotted line shows the result for galaxies with significant AGN and FIR photometry. The solid line shows the result for the entire DEVILS-D10 sample. \textit{Bottom:} The percentage of galaxies hosting AGN in the DEVILS-D10 data set. AGN are identified as galaxies with an AGN fraction > 0.1. We only show this result for galaxies that have FIR photometry otherwise the AGN component of the SEDs can become unconstrained. }
    \label{fig:agn_frac}
\end{figure}

The top panel of \Cref{fig:agn_frac} shows the median AGN fraction of galaxies in the DEVILS-D10 sample, where AGN fraction is defined as the ratio of the AGN flux between 5 and 20 $\mu$m to the total SED flux. 
The median AGN fraction of the entire DEVILS-D10 sample is negligible at $0<$ look-back time/Gyr $<12.5$. We also show this result for galaxies with an AGN fraction $> 0.1$ that is the definition of significant AGN used by \citet{thorneDeepExtragalacticVIsible2022a}. We see that the median AGN fraction decreases from $\approx 0.6$ to $\approx 0.4$ between look-back time $12.5 \to 8$ Gyr, before rising back up and flattening to $\approx 0.6$ from look-back time $\approx 8$~Gyr to $\approx 4$~Gyr. If we include the additional constraint that the SEDs must also have FIR photometry, then the median AGN fraction at look-back time $<8$ Gyr does not experience the rise in AGN fraction that we see when loosening this FIR constraint, staying flat at $\approx 0.3 \to 0.4$. The bottom panel of \Cref{fig:agn_frac} shows the percentage of galaxies hosting a significant AGN in the DEVILS-D10 data set. We show this result only for the galaxies that have FIR photometry. We see that the percentage of galaxies hosting AGN rises from $\lesssim 1$ per cent at look-back time of 12.5 Gyr to a peak of $\approx 4$ per cent at $\approx$ 1.5 Gyr, before declining to $\approx 2$ per cent at look-back time $\to 0$ Gyr. The declining percentage of galaxies hosting a significant AGN with look-back time is somewhat inconsistent with previous studies that suggest that the probability of galaxies hosting AGN is greater in the early Universe than at more recent times \citep[e.g.,][]{aird_2018_agn_prob}. This supposed inconsistency is likely due to differences in the definitions of significant AGN. In \textsc{ProSpect}, a significant AGN defined as one whose contribution to SED from the AGN is more than 10 per cent. At high redshift, galaxies are generally more star forming meaning that the SED is dominated by the emission from stars. This results in a low AGN fraction, and thus insignificant AGN according to these definitions, despite these AGN being potentially bolometrically luminous. Indeed, \Cref{fig:agn_density} shows that the peak of AGN density shifts $\approx 2$~dex toward higher AGN bolometric luminosities from $z \approx 0 \to 5$.

The sharp increase in the median AGN fraction between look-back time $\approx 8$ and $\approx4$~Gyr for galaxies whose SEDs are missing FIR photometry (the dashed line in the top panel of \Cref{fig:agn_frac}) is likely the origin of the deficit that we see in the CSFH in \Cref{fig:agn_no_agn}. As such the SFR replacements that we perform will have more of an impact on the CSFH at these later look-back times. The inclusion of an AGN component in the SED fitting has a greater effect on the SFR at lower look-back time compared to the early Universe because the average SFR is lower. This means that small changes in the MIR fit, to account for the AGN, can induce larger fractional changes in the SFR, especially in the cases where the AGN component is not constrained by FIR photometry. On the other hand, the average SFRs of galaxies are much higher in the early Universe than at later times and so the inclusion of a bright AGN component does little to drive down the net SFR of galaxies. This also explains the declining trend with look-back time seen in the bottom panel of \Cref{fig:agn_frac}.

\subsection{Cosmic star formation and AGN bolometric luminosity history}
\label{subsect:csfrd_cagn}

\begin{figure*}
    \centering
    \includegraphics[width = \textwidth]{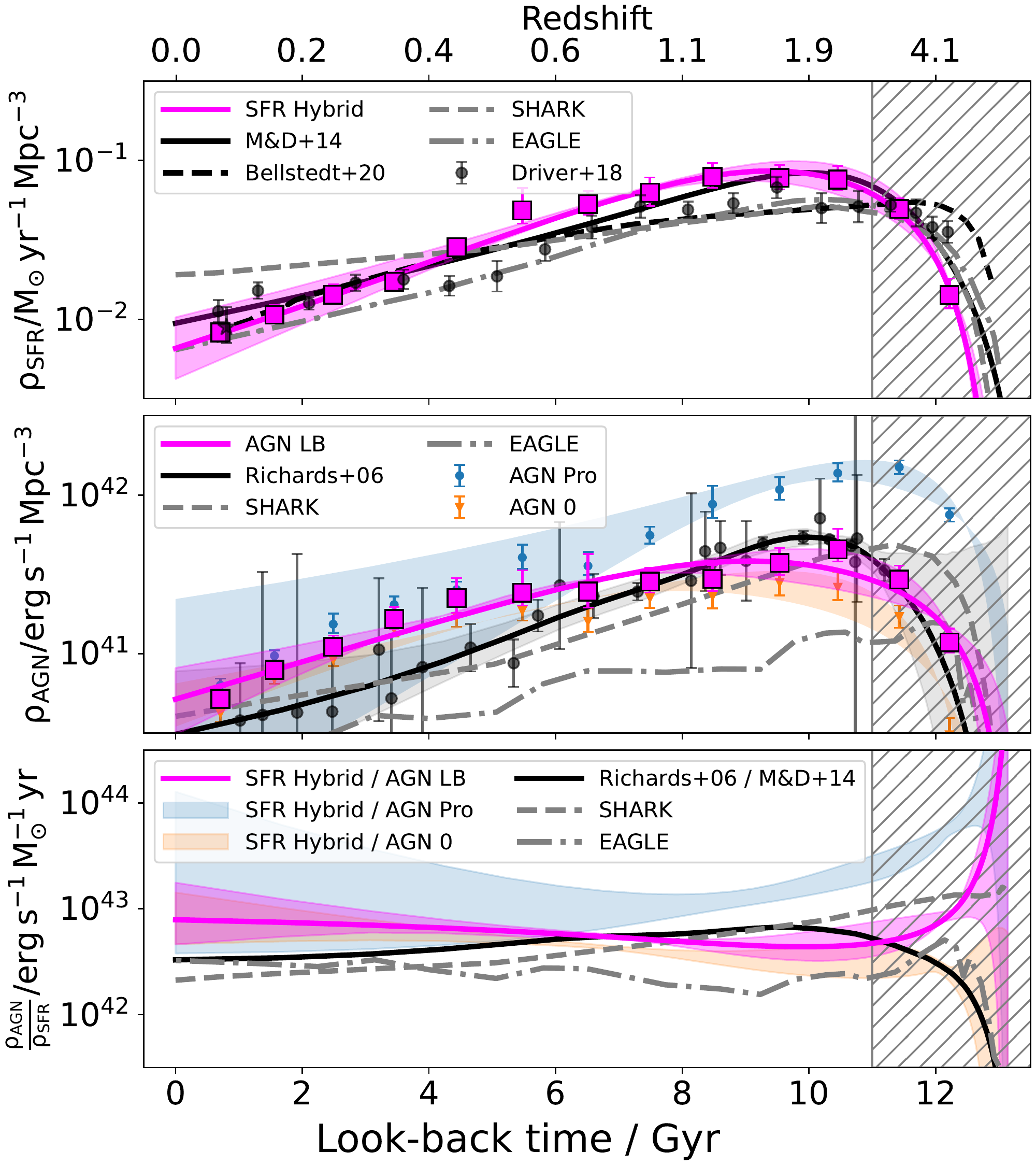}
    \caption{\textit{Top:} CSFH. The magenta points and line shows our results for the SFR$_{\mathrm{Hybrid}}$ sample that we discuss in \Cref{subsect:replace} and \Cref{tab:replace_defn}. Black circles show the results from \citet{driverGAMAG10COSMOS3DHST2018}. The black, star shaped point and dashed, black line shows the results from \citet{bellstedt_prospect_2020}. The solid, black line shows the result from \citet{madauCosmicStarFormationHistory2014}. The grey, dashed line shows the result from the  \textsc{Shark} semi-analytic model \citep{lagos_shark_2018}. The grey, dash-dotted line shows the results from the EAGLE reference model \citep{schaye_eagle_2015,crain_eagle_2015}. \textit{Middle:} CAGNH. The magenta points and line shows our results for the AGN$_{\mathrm{LB}}$ sample. The blue and orange points and shades show the results for the AGN$_{\mathrm{Pro}}$ and AGN$_{\mathrm{0}}$ samples. The black points and line shows the results from \citet{richards_sloan_2006}. The grey, dashed line shows the result from \textsc{Shark}. The grey, dot-dashed line shows the result from the EAGLE reference model. \textit{Bottom:} The ratio of cosmic AGN bolometric luminosity to SFR density. The magenta line shows the result for the SFR$_{\mathrm{Hybrid}}$ CSFH and the AGN$_{\mathrm{LB}}$ CAGNH. The blue and orange shaded regions show the results for the AGN$_{\mathrm{Pro}}$ and AGN$_{\mathrm{0}}$ samples, where for both we divide by the SFR$_{\mathrm{Hybrid}}$ CSFH. The black line shows the result for the \citet{richards_sloan_2006} CAGNH over the \citet{madauCosmicStarFormationHistory2014} CSFH. The grey, dashed line shows the result for \textsc{Shark}. The grey, dot-dashed line shows the result from the EAGLE reference model. In all of the panels we draw a hatched region at a look-back time $\geq 11$ Gyr that is the limit at which we consider these results reliable.}
    \label{fig:fits_master}
\end{figure*}

\subsubsection{CSFH and comparison with previous results}
The top panel of \Cref{fig:fits_master} shows our final CSFH for the hybrid sample of GAMA and DEVILS-D10 with the LSS correction of \Cref{fig:LSS} applied. We fit our data points with the \texttt{mass-func\_snorm\_trunc} function in redshift space (\Cref{eq:mass_func_snorm_trunc}) using the Markov-Chain-Monte-Carlo (MCMC) sampler \textsc{EMCEE} \citep{emcee}. We fix \texttt{magemax} to $z=20$ and \texttt{mtrunc}=1. We use uniform priors (U) on the normalisation, \texttt{mSFR}$\sim U(0,1)$; the peak, \texttt{mpeak}$\sim U(0.5,4.0)$; the width of the underlying Normal distribution, \texttt{mperiod}$\sim U(0.01,2.01)$; and the skewness of the Normal distribution, \texttt{mperiod}$\sim U(-1.0, 1.0)$. The fit parameters are presented in \Cref{tab:fits}.

\begin{table}
\begin{center}
\begin{tabular}{ |c|c|c|c|c| }
\hline
          - &        Parameter &    Fit &  $\sigma_{16}$ &  $\sigma_{84}$ \\
\hline
\textbf{SFR$_{\mathrm{Hybrid}}$} &    \texttt{mSFR} &  0.088 &          0.007 &          0.008 \\
         - &   \texttt{mpeak} &  1.581 &          0.108 &          0.114 \\
         - & \texttt{mperiod} &  1.015 &          0.051 &          0.051 \\
         - &   \texttt{mskew} & -0.299 &          0.047 &          0.044 \\
\hline
AGN$_{\mathrm{Pro}}$ &    $\log_{10}( \texttt{mAGN})$ & 42.171 &          0.037 &          0.035 \\
      - &   \texttt{mpeak} &  2.320 &          0.113 &          0.118 \\
      - & \texttt{mperiod} &  1.431 &          0.083 &          0.120 \\
      - &   \texttt{mskew} & -0.347 &          0.087 &          0.067 \\
\hline
\textbf{AGN$_{\mathrm{LB}}$} &    $\log_{10}(\texttt{mAGN})$ & 41.594 &          0.046 &          0.047 \\
                 - &   \texttt{mpeak} &  1.420 &          0.236 &          0.247 \\
                 - & \texttt{mperiod} &  1.098 &          0.136 &          0.125 \\
                 - &   \texttt{mskew} & -0.408 &          0.101 &          0.098 \\
\hline
AGN$_{\mathrm{0}}$ &    $\log_{10}(\texttt{mAGN})$ & 41.470 &          0.038 &          0.037 \\
                - &   \texttt{mpeak} &  1.334 &          0.162 &          0.170 \\
                - & \texttt{mperiod} &  0.971 &          0.079 &          0.078 \\
                - &   \texttt{mskew} & -0.290 &          0.072 &          0.069 \\
\hline
\end{tabular}
\end{center}
\caption{\texttt{mass-func\_snorm\_trunc} fit parameters to the CSFH and CAGNH. The sample names are the same as those described in \Cref{subsect:replace} and \Cref{tab:replace_defn}.}
\label{tab:fits}
\end{table}

A highlight of this figure is the remarkable agreement with the fit of the CSFH from \citet{madauCosmicStarFormationHistory2014} \footnote{The \citet{madauCosmicStarFormationHistory2014} result uses data from \citet{wyder2005,schiminovich2005,robothamdriver2011,cucciati2012,dahlen2007,reddysteidel2009,schenker2013,sanders2003,takeuchi2003,magnelli2011,magnelli2013,gruppioni2013}.}, which is encouraging considering that they use a swath of independent literature results that span the breadth of the electromagnetic spectrum to derive the CSFH. 

We also agree with the $z \approx 0.06$ and the forensic reconstruction (found by stacking the star formation histories of individual galaxies) of the CSFH up to look-back time of $\approx 4.5$ Gyr from \citet{bellstedt_prospect_2020} who use both GAMA photometry and \textsc{ProSpect} with identical parameters used in this work, and so is an appropriate data set for verification of our results. We note that the results presented in the main body of \citet{bellstedt_prospect_2020} used a closed-box metallicity evolution while here we show their results using the linear metallicity mapping (their appendix B) to be consistent with the metallicity mapping that was used to fit the SEDs of galaxies in this work. \citet{bellstedt_prospect_2020} do not account for an AGN component either when fitting the SEDs of their $z<0.06$ galaxies. 

We also exhibit broad agreement with the results of \citet{driverGAMAG10COSMOS3DHST2018} who use a similar method of spline fitting the SFR distributions. We expect that the differences between this work and \citet{driverGAMAG10COSMOS3DHST2018} can be pinned down to underlying differences in the photometry extraction (\textsc{SourceExtractor} against \textsc{ProFound}), SED fitting (\textsc{MagPhys} against \textsc{ProSpect}) and how they filter AGN (where the SFR contribution from AGN hosts is zero and the AGN fraction is effectively either 0 or 1). 

We also show the results from the \textsc{Shark} semi-analytic model \citep{lagos_shark_2018}, where there is tension with this work. \textsc{Shark} predicts a $\approx 0.3$ dex lower peak around look-back time $\approx 9$ Gyr than this work, which agrees more closely with the results of \cite{bellstedt_prospect_2020} and \citet{driverGAMAG10COSMOS3DHST2018}. This is unsurprising since the \citet{driverGAMAG10COSMOS3DHST2018} results were used as a consistency check for \textsc{Shark} \citep[``secondary observational constraint'' in the jargon of][]{lacey_galform_2016}. \textsc{Shark} predicts $\approx 0.3$ dex higher SFR density at look-back time 0 Gyr compared to our results and the literature results. The overestimate of star formation in \textsc{Shark} compared to this work may reflect that AGN feedback in \textsc{Shark} is not efficient enough at inhibiting star formation at low look-back times; though, we also note that \textsc{Shark} produces star formation in massive galaxies, $\mathrm{M_{\star} \gtrsim 10^{11}M_{\odot}}$, that are not well sampled in GAMA. A closer look at the physical models of star formation and feedback in \textsc{Shark} is beyond the scope of this work but will be the focus of forthcoming studies (Lagos et al. in prep; Bravo et al. in prep). 


We also compare our results to the Ref-L100 EAGLE cosmological, hydrodynamical simulation \citep{schaye_eagle_2015,crain_eagle_2015}. The EAGLE CSFH has a similar shape as our results at look-back time $\lesssim 11$~Gyr. EAGLE however predicts a much lower CSFH, with the biggest difference being $\approx 0.2$ dex lower than our result at a look-back time of $\approx 5$ Gyr. The $100$Mpc$^{3}$ volume of the EAGLE reference box can explain this deficit where the contribution to the cosmic SFR density from highly star forming galaxies is missing in EAGLE as they are not present in the simulation box. The tension could also hint that star formation is not efficient enough in the simulations, possibly as a result of the feedback implementation in the model. By look-back time 0 Gyr, the EAGLE result agrees with our work. 

\subsubsection{CAGNH and comparison to previous results}
\label{subsubsect:cagn}
The middle panel of \Cref{fig:fits_master} shows the estimated CAGNH for our three different samples of AGN bolometric luminosity (AGN$_{\mathrm{Pro}}$, AGN$_{\mathrm{LB}}$, AGN$_{\mathrm{0}}$) as per their descriptions in \Cref{subsect:replace} and \Cref{tab:replace_defn}. We note that these results also have had the LSS correction applied, and that the contribution from the additional literature datasets discussed in \Cref{sect:agn_literature} is the same for each of the three samples. The purpose of these three different samples is that for galaxies missing FIR photometry, the AGN component that \textsc{ProSpect} estimates from fitting the SED (AGN$_{\mathrm{Pro}}$) can become unconstrained. Thus, if we are to use the luminosities straight from \textsc{ProSpect} they are likely to be upper limits. Setting these potentially unconstrained AGN bolometric luminosities instead to 0 (AGN$_{\mathrm{0}}$) represents the minimum possible CAGNH solution, and, with the upper limit of AGN$_{\mathrm{Pro}}$, we can constrain a bounded shape to the CAGNH. However, in setting these AGN bolometric luminosities to 0, we are excluding the possibility that the galaxies missing FIR photometry retain any AGN component. So to present a more physically meaningful replacement, which is straddled by the AGN$_{\mathrm{Pro}}$ and AGN$_{\mathrm{0}}$ results, we draw particular attention to the CAGNH resulting from the AGN$_{\mathrm{LB}}$ sample, where the AGN bolometric luminosities of the galaxies missing FIR photometry are instead replaced with the lower bound AGN bolometric luminosity from \textsc{ProSpect}. 

We fit these points with the  \texttt{mass-func\_snorm\_trunc} function, using the same priors as we did for the CSFH but this time adjusting the normalisation prior to be $\log_{10}$(\texttt{mAGN})$\sim U(38,48)$. Again, the fit parameters are presented in \Cref{tab:fits}. We find that the AGN$_{\mathrm{LB}}$ CAGNH follows the CSFH in the top panel of \Cref{fig:fits_master} closely. Interestingly, we find that the AGN$_{\mathrm{LB}}$ and AGN$_{\mathrm{0}}$ samples yield similar results, with the biggest difference being $\approx 0.7$ dex in the highest look-back time bin of >12.0 Gyr, but we note that the data are subject to the greatest selection effects in these high look-back time bins. 

For the AGN$_{\mathrm{Pro}}$ CAGNH, we find that the normalisation of the points is so high that the CAGNH remains fairly flat ($\approx 10^{42} \mathrm{erg \, s^{-1}Mpc^{-3}}$) at look-back time $\gtrsim 10$ Gyr. Thus, the \texttt{mass-func\_snorm\_trunc} function cannot properly capture the slopes of the CAGNH at both early and recent times. This, in turn, propagates through to a larger uncertainty, especially at look-back time $\lesssim 8$ Gyr. Furthermore, we observe in \Cref{fig:agn_no_agn} that the CSFH from the AGN included run of \textsc{ProSpect} is slightly lower than that of the SFR only run at look-back time $\gtrsim 8$ Gyr. All of which strengthen our suspicion that the AGN$_{\mathrm{Pro}}$ CAGNH is an upper limit. 

We compare these results with \citet{richards_sloan_2006} who use the Sloan Digital Sky Survey (SDSS) to compute the analytic quasar luminosity function that we then use to compute the CAGNH by integration of the fitted light functions. We fit these points with an eighth order smooth spline, weighting by the inverse of the variance. An error bound on these fits is constrained by refitting 1001 times and jostling the data points within their uncertainties in each iteration. Such a fitting scheme was found to best represent the data, especially at look-back time $\to 0$ where there are large uncertainties. There is overall good agreement between \citet{richards_sloan_2006} and the AGN$_{\mathrm{LB}}$ and AGN$_{\mathrm{0}}$ samples at look-back time $\gtrsim 6$ Gyr. At lower look-back times, we appear to estimate a shallower decline of AGN activity compared to \cite{richards_sloan_2006}, potentially indicative of us recovering more AGN than \citet{richards_sloan_2006}. Though the large uncertainties in the \citet{richards_sloan_2006} results prevent a robust conclusion. 

We find that the \textsc{Shark} curve peaks earlier ($\approx 10$ Gyr compared to $\approx 11$ Gyr ago) and more sharply than ours as evidenced by the narrower width of the peak. Our results also exhibit a shallower decline from $\approx 10$ Gyr ago compared to \textsc{Shark}. The EAGLE curve also peaks earlier and more sharply than our results, but experiences a similar decline as our results. We also see that the EAGLE CAGNH is lower in normalisation to ours that can be explained by the small volume of the EAGLE reference box where not enough bright AGN are sampled. The EAGLE curve is also noisier than our results that can be explained by cosmic variance, again as a consequence of the small volume. 

\subsubsection{Interface of CSFH and CAGNH}
The bottom panel of \Cref{fig:fits_master} shows the ratio of the CAGNH to the CSFH, where the CSFH is only from the SFR$_{\mathrm{Hybrid}}$ sample. In the presentation of this figure, we wish to show the shapes of the CSFH and CAGNH in relation to one another, allowing us to see which processes are dominant over time. The AGN$_{\mathrm{Pro}}$ CAGNH shows a declining trend up to look-back time $\approx 8$~Gyr where we would infer that the evolution of AGN activity is dominant to SFR. At look-back time $\lesssim 8$~Gyr the ratio becomes too unconstrained, as a result of the large uncertainty on the AGN$_{\mathrm{Pro}}$ CAGNH, to make meaningful conclusions on the time evolution of the AGN$_{\mathrm{Pro}}$ upper limit. 

At look-back time $\gtrsim 11$ Gyr, we see vast differences between each of our AGN subsets, which should straddle a range of viable AGN solutions, highlighting that the first few billion years after the Big Bang are currently unconstrained by even our state-of-the-art multi-wavelength observations. As such, in all of the panels in \Cref{fig:fits_master} we mark with a hatched region at look-back time $\gtrsim 11$ Gyr the point beyond which the results are likely to be unreliable. Understanding the intricacies of early star and black hole formation in the first couple of billion years after the Big Bang will thus be a task for high-redshift instruments like the Hubble Space telescope, \textit{JWST} and the Atacama Large Millimetre Array, leveraging a wide range of wavelengths for full SED analyses. 

At look-back time $\lesssim 11$ Gyr, the lines for both the AGN$_{\mathrm{LB}}$ and AGN$_{\mathrm{0}}$ samples are fairly constant ($\approx 10^{42.5} \mathrm{erg \, s^{-1}M_{\odot}^{-1}yr}$) with low scatter ($\lesssim 0.3$ dex), suggesting that star formation and AGN activity are coeval. Moreover, the AGN$_{\mathrm{Pro}}$ result also exhibits little evolution at look-back time $\lesssim 8$ Gyr (albeit with much larger scatter). Remarkably the ratio of CAGNH to CSFH are all within $\approx 0.5$ dex at look-back time $\lesssim 11$ Gyr, regardless of which AGN bolometric luminosities we use, highlighting the strength of this result. 

If we assume that the AGN bolometric luminosity is driven by accretion with a radiative efficiency of 10 per cent, i.e., $\mathrm{L_{AGN}} = \dot{\mathrm{M}}_{\mathrm{BH}}\times \epsilon c^{2}$ where $\mathrm{L_{AGN}}$ is the AGN bolometric luminosity, $\dot{\mathrm{M}}_{\mathrm{BH}}$ is the black hole accretion rate and $c$ is the speed of light in a vacuum, then the dimensionless $\mathrm{CAGNH / CSFH} \approx 5.6\times10^{-4}$ at look-back time $\lesssim 11$ Gyr. If we compare this to the average black hole mass to stellar mass ratio, $\langle \mathrm{M_{BH}/M_{\star}} \rangle \approx 10^{-3}$ \citep[e.g.,][]{Kormendy_Ho_smbh_2013}, then we get $5.6\times10^{-4} / 10^{-3} \approx 0.5$.

Stellar evolution dictates that a fraction of the mass in stars is gradually returned to the interstellar medium (ISM) implying that the correspondence between star formation rate and stellar mass is not exact. As such, we can explain the offset between $\mathrm{CAGNH/CSFH}$ and $\langle \mathrm{M_{BH}/M_{\star}} \rangle$ if $\approx 50$ percent of the mass in stars is returned to the ISM by $z=0$. Interestingly, this is precisely the return fraction assumed in \Cref{subsect:LSS} when we calculated the cosmic stellar mass density for our large scale structure correction, pointing to a potential consistency.


For comparison, we calculate the ratio of our fit to the \citet{richards_sloan_2006} CAGNH and the fit to the CSFH by \citet{madauCosmicStarFormationHistory2014}. This line shows a steeper decline of AGN activity compared to star formation, contrary to our finding; though, we remark that the large uncertainty in the \citet{richards_sloan_2006} CAGNH will propagate through to a large uncertainty in the ratio of CAGNH and CSFH. Now turning to the simulations, in \textsc{Shark} we find that the ratio of CAGNH to CSFH favours the CAGNH suggesting that the rising slope of the CAGNH in the early Universe is steeper to that of the CSFH. In EAGLE the opposite trend is observed as the ratio favours the CSFH compared to the CAGHN up until a look-back time of $\approx 12$ Gyr. In both of the simulations, the slope of the ratio favours more so the SFR at look-back time $\lesssim 10$ Gyr, in tension with our results that suggest these two processes are actually coeval.

\subsubsection{General interpretations}
The fact that we see little evolution of the ratio of CAGNH to CSFH is interesting given that in low to moderate power AGN hosting galaxies ($L_{X} \lesssim 10^{44} \, \mathrm{erg \, s^{-1}}$) there is a weak correlation between star formation and black hole accretion rate \citep[e.g.,][]{shao_starformation_2010,mullaney_goods_2012,rosario_meanstarformation_2012,harrison_noclear_2012, azadi_primus_2015}. The weak correlation in individual galaxies can be explained by the variability of AGN luminosities over short timescales ($\approx 10^{6}-10^{7}$ Myr) compared to star formation ($\approx 10^{8}$ Myr) because the relaxed phase of the AGN duty cycle occurs more often than a relaxed phase of star formation \citep{hickox_black_2014}. Thus, the chance that we would observe AGN that are ``switched on'' compared to actively star forming galaxies is rarer and the ratio of CAGNH to CSFH would instead favour star formation. Instead, the flat evolution that we see in our results indicates that short timescale variations in AGN luminosity average out in large samples, such as the combined set of GAMA and DEVILS-D10 we use in this work. Indeed, AGN activity has previously been shown to correlate with star formation activity when considering global galaxy populations \citep{boyle_cosmological_1998}. 


Ultimately, the flat CAGNH/CSFH indicates that the mechanism behind the coeval rise and fall of SF and AGN activity over 11 Gyr is of the same origin and operates at the same rate.

\begin{figure}
    \centering
    \includegraphics[width=\linewidth]{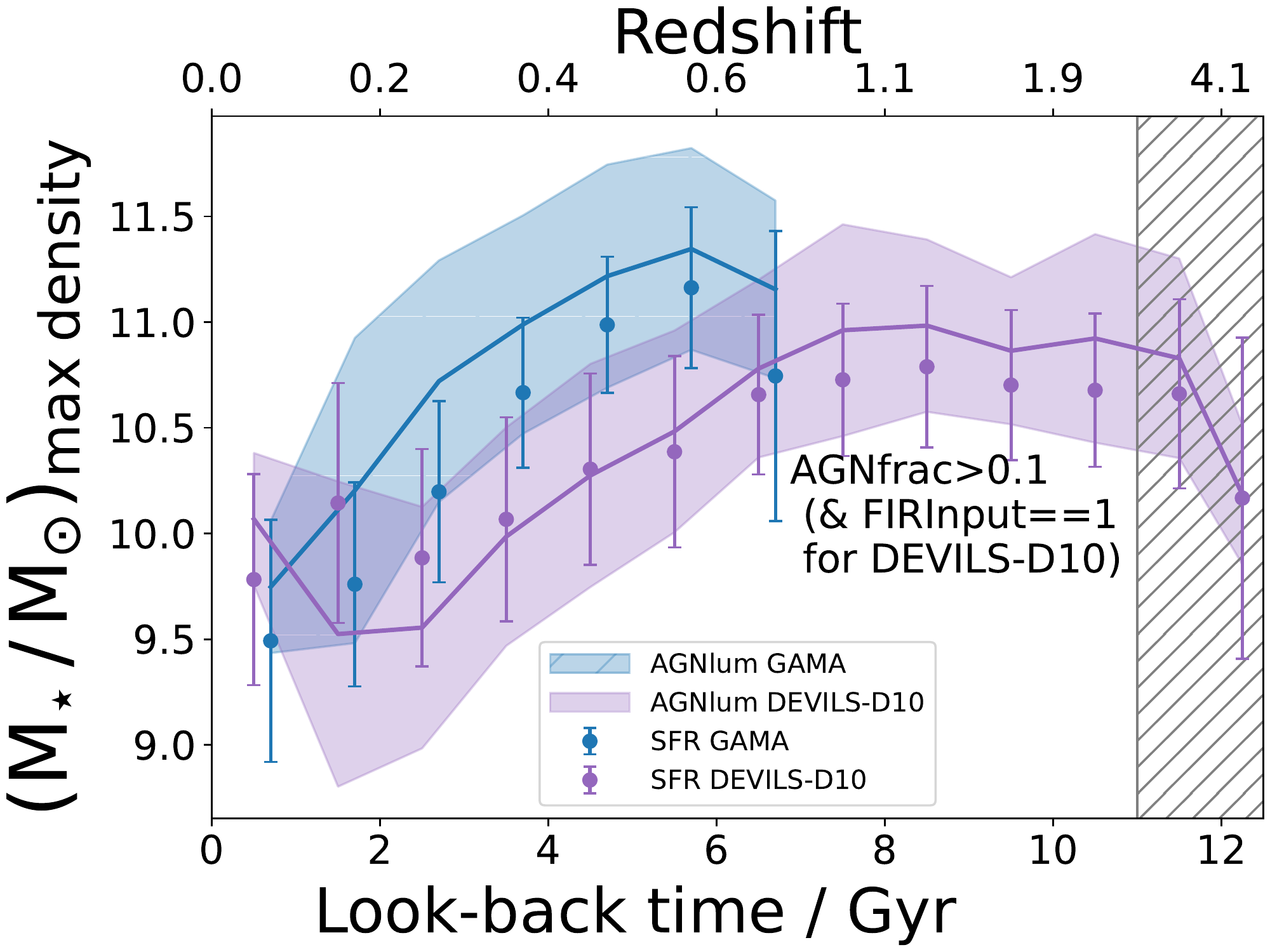}
    \caption{Median and $1\sigma$ spread of stellar mass of galaxies that contribute at least 50 per cent of the cosmic SFR (circles and error bars) and AGN bolometric luminosity (lines and shaded regions) density per unit look-back time in GAMA (blue) and DEVILS-D10 (purple). For clarity, the GAMA points have been shifted slightly to the right. We find the SFR and AGN bolometric luminosity boundary values to the left and right of the peak of the density distributions of \Cref{fig:sfr_density} and \Cref{fig:agn_density} such that the integrals within those boundaries are at least 50 percent the value of the total cosmic SFR/ AGN bolometric luminosity density. We use the AGN$_{\mathrm{LB}}$ results for our AGN selection. We only show this result for confirmed AGN with an AGN fraction greater than 0.1 and with FIR photometry present in the galaxy SEDs.}
    \label{fig:median_stellar_mass}
\end{figure}

To further investigate this idea, we calculate the median and $1\sigma$ spread of stellar mass of the galaxy sample that lie within the peak of the distributions in \Cref{fig:sfr_density} and \Cref{fig:agn_density}. We find the boundaries to the left and right of the maximum of either the SFR or AGN bolometric luminosity density distributions such that the integral within those boundaries is at least 50 per cent of the total area under the curves. We then select galaxies with SFRs or AGN bolometric luminosities within those boundary values and calculate the median and 1$\sigma$ spread of stellar mass in that SFR/AGN bolometric luminosity bin. We only show this result for confirmed AGN where the AGN fraction is $>0.1$ and the SEDs have FIR points for the DEVILS-D10 sample. The result of this exercise is shown in \Cref{fig:median_stellar_mass}. 

There is a slight tendency for more massive galaxies to dominate the CAGNH compared to the CSFH in GAMA only, but both are still well within their respective $1\sigma$ spreads of stellar mass meaning we cannot confidently determine if there is a true difference in these two populations. Considering both GAMA and DEVILS, we find that the galaxies with significant AGN contributing to at least half of the cosmic SFR density are similar to the ones contributing to the cosmic AGN bolometric luminosity density with stellar masses within 1$\sigma$, indicating that star formation and AGN activity are linked. The link likely rests in the gas supply in galaxies being the same fuel for both star formation and AGN activity. So, as the gas supply is continuously consumed star formation and AGN activity must reduce equivalently to ensure that the slopes of their declines at look-back time $\lesssim 10$~Gyr are similar.

\section{Conclusions}
\label{sect:conclusion}
We have calculated the joint evolution of the CSFH and CAGNH over 12.5 Gyr since the present day by using SED fits to galaxies in the GAMA and DEVILS-D10 data sets. \textsc{ProSpect} was used to fit the SEDs, simultaneously and self-consistently modelling the flux associated with stars and AGN. The key results are summarised below.

\begin{itemize}
    \item After carefully accounting for the potential underestimation of star formation due to the presence of an AGN component in the SEDs and the effect of large scale structure, our CSFH agrees well with literature results \citep[e.g.,][]{madauCosmicStarFormationHistory2014, driverGAMAG10COSMOS3DHST2018, bellstedt_prospect_2020}, with a minimum at look-back time $\approx 12.5$ Gyr, a peak at $\approx 10$ Gyr and a decline into the present day. We expect slight differences between our results and the literature to be driven by differences in the photometry extraction, SED fitting method and AGN selection used in the literature. 
    \item Our CAGNH follows a similar shape to the CSFH. We report a shallower decline of AGN bolometric luminosity density at look-back time $\lesssim 8$ Gyr than \citet{richards_sloan_2006}. 
    \item Taking the ratio of the CAGNH and the CSFH, we find that AGN activity and star formation have been coeval, with a ratio of $\mathrm{\rho_{AGN}/\rho_{SFR}}\approx 10^{42.5} \mathrm{erg \, s^{-1}M_{\odot}^{-1}yr}$, for the last $\approx 11$ Gyr. We find that whether we use the AGN bolometric luminosities straight out of \textsc{ProSpect} or replace them with either 0 or the lower bound AGN luminosity, the AGN bolometric luminosity density is within 0.5 dex over the last $\approx 11$ Gyr.  
    \item At look-back time $\gtrsim 11$ Gyr, the CAGNH and CSFH are unconstrainable with these state-of-the-art multiwavelength datasets. It will be a task for high-redshift instruments like the  Hubble Space Telescope and \textit{JWST} to better constrain the AGN activity at look-back time $\gtrsim 12.5$~Gyr ($z \gtrsim 5$). 
    \item We show that the CSFH and CAGNH in the semi-analytic model \textsc{Shark} and the cosmological hydrodynamical model EAGLE exhibit slight differences with our results. For example, \textsc{Shark} produces $\approx 0.3$ dex more cosmic SFR density at look-back time $\to 0$ Gyr compared to our result, while EAGLE produces $\approx 0.2$ dex less cosmic SFR density at look-back time $\approx 5$ Gyr. Both \textsc{Shark} and EAGLE predict a sharper decline of AGN activity compared to SFR than our results. The key point is that this work will serve to inform the physical models of star formation and AGN feedback in simulations. 
\end{itemize}

As galaxies are multifaceted in nature we must go beyond isolated considerations of either star forming galaxies or galaxies hosting AGN. Thanks to the exquisite GAMA and DEVILS-D10 photometry and \textsc{ProSpect} SED fits, this work represents an important milestone toward reconciling the role of star formation and AGN activity in the life cycle of galaxies over $\approx 12.5$ Gyr from the present day.

\section*{Acknowledgements}
We thank the anonymous referee for their suggestions and advice that improved the quality of this work. 

JCJD is supported by the Australian Government Research Training Program (RTP) Scholarship. 

DEVILS is an Australian spectroscopic campaign that uses the Anglo-Australian Telescope. DEVILS is part funded via Discovery Programs by the Australian Research Council and the participating institutions. The DEVILS input catalogue is generated from data taken as part of the ESO VISTA- VIDEO \citep{jarvis_vista_2013} and UltraVISTA \citep{mccracken_ultravista_2012} surveys. The DEVILS website is \url{https://devilsurvey.org}. The DEVILS data is hosted and provided by AAO Data Central (\url{https://datacentral.org.au/}).

GAMA is a joint European-Australasian spectroscopic campaign using the Anglo-Australian Telescope. The GAMA input catalogue is based on data taken from the Sloan Digital Sky Survey and the UKIRT Infrared Deep Sky Survey. Complementary imaging of the GAMA regions is being obtained by a number of independent survey programmes including GALEX MIS, VST KiDS, VISTA VIKING, WISE, \textit{Herschel}-ATLAS, GMRT and ASKAP providing UV to radio coverage. GAMA is funded by the STFC (UK), the ARC (Australia), the AAO, and the participating institutions. The GAMA website is \url{http://www.gama-survey.org/}. Based on observations made with ESO Telescopes at the La Silla Paranal Observatory under programme IDs 177.A-3016, 177.A- 3017, 177.A-3018 and 179.A-2004.

This work was supported by resources provided by the Pawsey Supercomputing Centre with funding from the Australian Government and the Government of Western Australia. We gratefully acknowledge DUG Technology for their support and HPC services. MB is funded by McMaster University through the William and Caroline Herschel Fellowship. This work has been supported by the Polish National Agency for Academic Exchange (Bekker grant BPN/BEK/2021/1/00298/DEC/1), the European Union's Horizon 2020 Research and Innovation programme under the Maria Sklodowska-Curie grant agreement (No. 754510).


\section*{Data Availability}

 The DEVILS and GAMA data products described in this paper are currently available for internal team use for proprietary science and will be made available in upcoming data releases. Scripts used for analysis and visualisation in this work will be made available upon reasonable request to the corresponding author.



\bibliographystyle{mnras}
\bibliography{ref} 




\appendix




\bsp	
\label{lastpage}
\end{document}